\documentclass[aps,prl,reprint,amsmath,amssymb,superscriptaddress,nofootinbib,noeprint,longbibliography,floatfix,nobibnotes]{revtex4-2}
\pdfoutput=1
\usepackage{newtxtext}
\usepackage{newtxmath}
\usepackage{dsfont}
\usepackage{braket}
\usepackage{xcolor}
\usepackage{microtype}
\usepackage{placeins}
\definecolor{myciteColor}{rgb}{0.0,0.5,0.23}
\usepackage[colorlinks=true,citecolor=myciteColor,linkcolor=myciteColor,urlcolor=myciteColor]{hyperref}
\usepackage{orcidlink}

\begin{abstract} \normalfont We study the ground-state phase diagram of a one-dimensional $\mathbb{Z}_2$ lattice gauge theory coupled to soft-core bosonic matter at unit filling, inspired by the Higgs sector of the standard model. Through a combination of analytical perturbative approaches, exact diagonalization, and density-matrix-renormalization-group simulations, we uncover a rich phase diagram driven by gauge-field-mediated resonant pair hopping and the confinement of single particles. The pair hopping results in a bunching state with superextensive energy and macroscopic particle number fluctuations at strong electric field strengths and weak on-site interactions. The bunching state crosses over into a pair superfluid phase as the on-site interaction increases, characterized by a finite superfluid density and powerlaw-decaying pair correlations. At large on-site interaction strengths and driven by effective interactions induced by the gauge constraint, the superfluid transitions into an incompressible pair Mott insulator phase. At weak field strengths and on-site interactions, we find a plasma-like region, where single bosons exhibit large short-range correlations and the ground state is composed almost equally of states with even and odd local boson occupation. The presence of a bunching state with large number fluctuations, which is difficult to study using classical numerics, motivates experimental realizations in hybrid boson-qubit quantum simulation platforms such as circuit QED, neutral atoms, and trapped ions. Our findings  highlight the rich interplay between gauge fields and soft-core bosonic matter.
\end{abstract}

\begin{document}
\title{Constrained many-body phases in a \texorpdfstring{$\mathbb{Z}_2$-}{Z2 }Higgs lattice gauge theory}
\author{Alexander Schuckert\orcidlink{0000-0002-9969-7391}} 
\email{alexander@schuckert.org}
\affiliation{Joint Quantum Institute and Joint Center for Quantum Information and Computer Science,
University of Maryland and NIST, College Park, Maryland 20742, USA}
\author{Stefan Kühn\orcidlink{0000-0001-7693-350X}}
\affiliation{Deutsches Elektronen-Synchrotron DESY, Platanenallee 6, 15738 Zeuthen, Germany}
\author{Kevin C. Smith\orcidlink{0000-0002-2397-1518}}
\altaffiliation{Present address: IBM Quantum, Cambridge, MA}
\affiliation{Brookhaven National Laboratory, Upton, New York 11973, USA}
\affiliation{Yale Quantum Institute, PO Box 208 334, 17 Hillhouse Ave, New Haven, CT 06520-8263, USA}
\affiliation{Departments of Applied Physics and Physics, Yale University, New Haven, CT 06511, USA}
\author{Eleanor Crane\orcidlink{0000-0002-2752-6462}} 
\affiliation{Departments of Physics and Mechanical Engineering, Co-Design Center for Quantum Advantage, Massachusetts Institute of Technology, Cambridge, Massachusetts 02139, USA}
\author{Steven M. Girvin\orcidlink{0000-0002-6470-5494}}
\affiliation{Yale Quantum Institute, PO Box 208 334, 17 Hillhouse Ave, New Haven, CT 06520-8263, USA}
\affiliation{Departments of Applied Physics and Physics, Yale University, New Haven, CT 06511, USA}
\maketitle

\textbf{Introduction.} Lattice gauge theory~\cite{Wilson1974,Kogut1975} (LGT) forms one of the cornerstones of our understanding of fundamental physics as it gives numerical access to nuclear and particle physics phenomena in the strong coupling regime. However, solving the full standard model of particle physics with its coupled matter, gauge and Higgs field sectors remains challenging to date. To nevertheless make progress, simple phenomenological models that enable the study of some of its phenomena have been developed. One of these models is the $\mathbb{Z}_2$ LGT~\cite{Wegner1971,Horn1979}, which uses a single two-level system as the gauge degree of freedom. Despite its simplicity, the $\mathbb{Z}_2$ LGT coupled to matter in one spatial dimension shows signatures of confinement~\cite{Borla2020} in one spatial dimension, reminiscent of confinement of quarks in quantum chromodynamics. So far, most studies have focused on 1D $\mathbb{Z}_2$ LGT coupled to fermions~\cite{Schweizer2019,Barbiero2018,Borla2020,bazavan2023,Davoudi2023,bennewitz2024,de2024,surace2024,mildenberger2025} or 2D $\mathbb{Z}_2$ LGT coupled to spins~\cite{mueller2024,cochran2024,gonzalezcuadra2024,borla2025}. Studies on soft-core bosons coupled to gauge fields, relevant to the Higgs sector, have focused on $U(1)$ LGT~\cite{Chanda2020,Chanda2022a}, with studies on $\mathbb{Z}_2$ models limited to cases with explicitly broken gauge symmetry~\cite{gonzalezcuadra2019,chanda2022,watanabe2025}. Therefore, to the best of our knowledge, phenomenological models of gauge fields coupled to genuine (i.e., soft-core) bosonic matter have been left largely unexplored, despite their relevance and potential for insight toward the Higgs sector of the standard model. In addition, a motivation for such models arises in the context of the cuprates~\cite{Senthil2000} and due to the possibility of their study in qubit-boson quantum computing hardware~\cite{Marcos2013,davoudi2020,Belyansky2024,crane2024}.

\begin{figure}[t]
    \centering
    \includegraphics[width=\columnwidth]{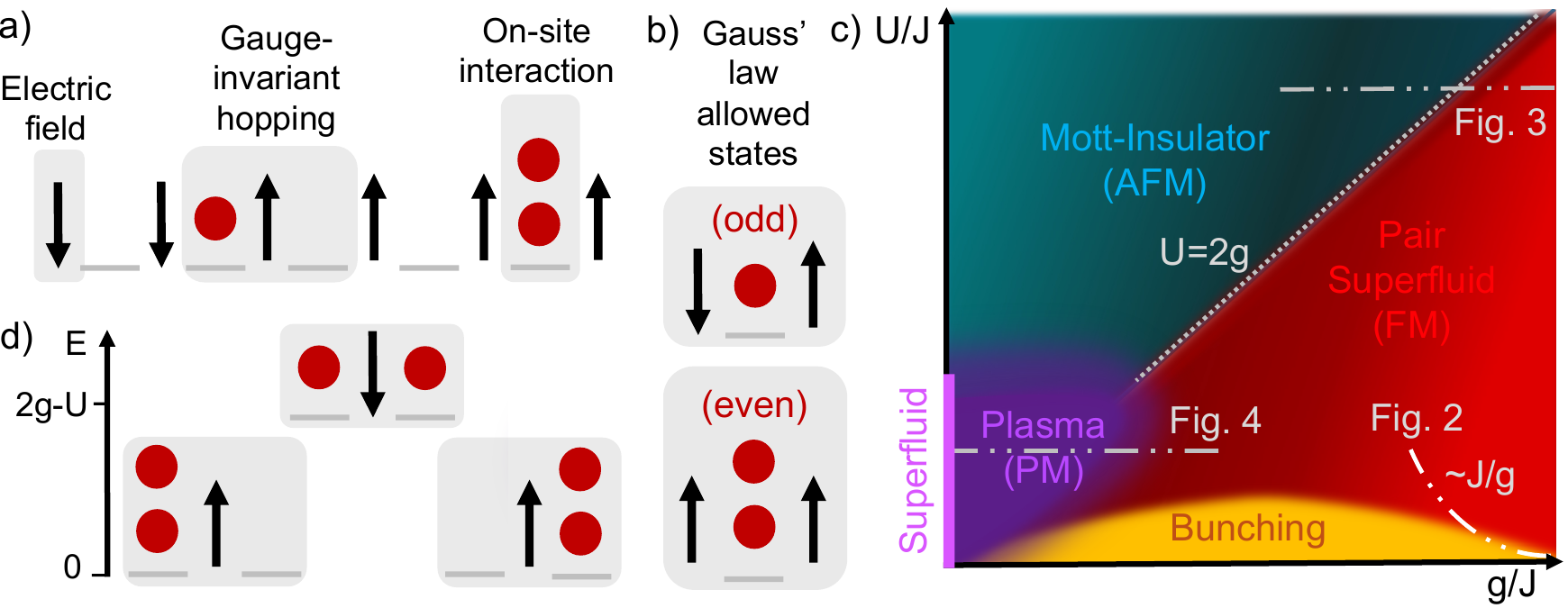}
    \caption{\textbf{The one-dimensional $\mathbb{Z}_2$-Higgs model.} a) Bosonic matter (red circles), residing on the sites, are coupled by the $\mathbb{Z}_2$ gauge field, represented by two-level systems on the links. The electric field term in the Hamiltonian acts on the gauge field on the links. The on-site interaction adds an energy penalty when two bosons occupy the same site. When a boson hops, it flips the gauge field on the link. b) Gauss' law is such that an odd (even) occupancy must have opposing (same) gauge fields to its left and right. c) Phase diagram at unit filling revealed by this work. The phase transition at $g/J=0, U/J\approx 3.45$~\cite{Elstner1999} is that of the usual Bose-Hubbard model at unit filling. The transition between Mott insulator and pair superfluid is at $U/J\approx 2 g/J$. The gauge field state is directly determined by the boson state through Gauss' law: the Mott-insulator, pair superfluid, plasma corresponds to classical Néel anti-ferromagnet (AFM), ferromagnet (FM) and paramagnet (PM), respectively. d) Energy diagram for the perturbative pair-hopping process. }
    \label{fig:1} 
\end{figure}

\textbf{Model.} In this work, we study the ground-state phase diagram of a $\mathbb{Z}_2$ LGT coupled to soft-core bosonic matter using a combination of the density matrix renormalization group~\cite{Schollwoeck2011, Orus2014} (DMRG), exact diagonalization, and analytical methods. Inspired by the Higgs sector of the standard model, we study the one-dimensional Hamiltonian 
\begin{equation}
    \hat{H} = - g \sum_{i=1}^{L-1} \hat{Z}_{{i,i+1}} - J \sum_{i=1}^{L-1}\left( \hat{a}^\dagger_{i} \hat{X}_{{i,i+1}} \hat{a}_{i+1} + \mathrm{h.c.} \right) + \frac{U}{2} \sum_{i=1}^{L} \hat{n}_i^2
    \label{eq_Z2_oneline}
\end{equation} 
at unit filling (one boson per site), where $\hat{a}_i$ and $\hat{a}_i^{\dagger}$ are the creation and annihilation operators of bosons fulfilling $[\hat{a}_i,\hat{a}^\dagger_j]=\delta_{ij}$, $\hat{n}=\hat a^\dagger \hat a$ is the bosonic number operator, and $\hat{X}_{i,i+1}$ and $\hat{Z}_{i,i+1}$ are Pauli operators of the $\mathbb{Z}_2$ gauge field sites linking two matter sites, see Fig.~\ref{fig:1}a. In analogy to the Higgs sector, the first, second, and third terms represent the electric field energy, the gauge invariant kinetic energy, and the local self-interaction of the bosonic Higgs field, respectively. In addition to bosonic particle number conservation, this Hamiltonian possesses a local gauge symmetry which we fix to
\begin{equation}
    \hat Z_{i-1,i} (-1)^{\hat n_i} \hat Z_{i,i+1}=1\label{eq:Gauss}
\end{equation}
in analogy to the physical Gauss' law constraint in electrodynamics in the absence of background charges, see Fig.~\ref{fig:1}b. While the gauge fields can in principle be integrated out~\cite{Pardo2023}, schemes to integrate out the bosons in the hard-core limit~\cite{Borla2020,Kebric2021} cannot be directly used here due to the many-to-one relationship between the boson occupation and the parity. This fact serves as a first indicator of the relevance of the bosonic nature of the matter fields. As we will show in the following, the  possibility of many bosons occupying the same site can increase the constraint on the Hilbert space induced by Gauss' law and subsequently lead to an intricate phase diagram, see Fig.~\ref{fig:1}c.

\textbf{Single-particle confinement}. A first natural question to ask is how the gauge-invariant single-particle correlator decays with distance:
\begin{equation}
    C^{(1)}_{ij}=\braket{\hat a^\dagger_i \prod_{k=i}^{j-1}\hat X_{k,k+1}\hat a_j+\mathrm{h.c.}},
\end{equation}
which serves as the order parameter for the confinement-deconfinement transition, where confinement (deconfinement) is indicated by an exponential (power-law) decay of $C^{(1)}_{ij}$ with respect to $|i-j|$~\cite{Borla2020}. For $g/J=0$, the model is equivalent to the 1D Bose-Hubbard model as the field sites decouple from the bosons. Consequently, $C^{(1)}$ becomes equal to the off-diagonal-long-range-order parameter of the superfluid-to-Mott-insulator phase transition which occurs at $U/J=3.45$~\cite{Elstner1999}. In the Mott-insulating phase, $C^{(1)}_{ij}$ decays as an exponential, while in the superfluid phase it decays polynomially with distance. For $g>0$, a single particle hopping from one site to the next involves flipping a field, incurring a cost of $2g$ in field energy. Hence, even for $g\ll J$, one would expect hopping of a particle over distance $r$ to be exponentially suppressed in $r$ as it involves a perturbative hopping process of amplitude $\frac{J^{r+1}}{2g^r}$. Indeed, we numerically find from our DMRG simulations that $C^{(1)}_{ij}$ decays exponentially for long distances for all $g>0$ studied, indicating that single particles are confined for all non-zero field strengths, analogous to the hard-core boson case~\cite{Kebric2021}. While from this observation one might expect no change in qualitative behavior of the system between soft- and hard-core bosons, this is not the case: as we will now describe, pairs of bosons display drastically distinct behavior from single bosons in the present model.

\textbf{Two-particle effective Hamiltonian.} Indeed, even for $g>0$ one expects pairs of bosons to be able to hop resonantly as such a process does not change the field energy, c.f.\ the superexchange~\cite{macdonald1988}-like energy diagram in Fig.~\ref{fig:1}d. This behavior is made most explicit in the regime $g\gg J\gg U$, where we find the effective Hamiltonian
\begin{align}
	\hat H \approx &-g \sum_i \hat Z_{i,i+1} -J_\mathrm{pair}\sum_i \hat Z_{i,i+1}\left(\left(\hat a_i^\dagger \right)^2  \left(\hat a_{i+1}\right)^2 + \mathrm{h.c.} \right)\notag \\&-	2J_\mathrm{pair}\sum_i\hat Z_{i,i+1}\hat n_i (\hat n_{i+1}+1)   + U \sum_i \hat n_i^2+\mathcal{O}\left(\frac{J^3}{g^2}\right)\label{eq:effH}
\end{align}
with pair hopping strength $J_\mathrm{pair}=\frac{J^2}{2g}$ via a Schrieffer-Wolff transformation (see End Matter). In this limit, bosons therefore assume a pair hopping as well as a nearest-neighbor attraction. Furthermore, the first term in the Hamiltonian forces all fields to be in $\ket{\uparrow}$, which we may interpret as a constraint on the ground space, $\hat Z_{i,i+1}=1$ for all $i$. While the effective Hamiltonian does not directly couple fields and bosons, this constraint on the fields has a direct consequence on the bosons through simplifying the Gauss' law constraint in Eq.~\eqref{eq:Gauss} to $(-1)^{\hat n_i}=1$. Therefore, in this limit, only states with an even number of bosons are part of the ground space.

\begin{figure}[t]
    \centering
    \includegraphics{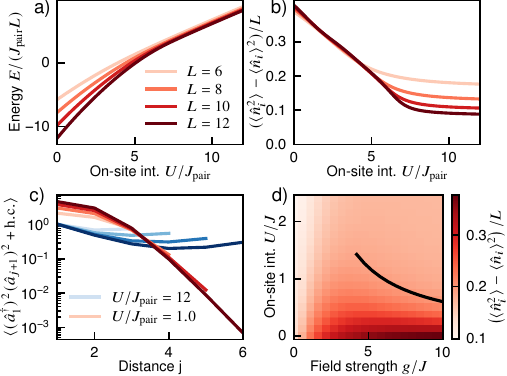}
    \caption{\textbf{Bunching region.} Results from exact diagonalization of (a-c) the perturbative Hamiltonian in Eq.~\eqref{eq:effH}, (d) the full Hamiltonian in Eq.~\eqref{eq_Z2_oneline}. Periodic boundary conditions. a) Energy density as a function of the on-site interactions strength. b) On-site fluctuations divided by the system size, as a function of on-site interaction strength. System sizes same as in a). c) Pair correlator as a function of distance (to reduce finite-size effects, we only show distances up to the middle of the system), within ($U/J_\mathrm{pair}=1$, red) and outside ($U/J_\mathrm{pair}=12$, blue) of the bunching region. System sizes same as in a).  d) On-site fluctuations in the full Hamiltonian for $L=6$ as a function of on-site interaction and field strength. The solid line is $U_\mathrm{c}/J=6(g/J)^{-1}$, the estimated boundary of the bunching region for large $g/J$.
    \label{fig:clump}}
\end{figure}

\textbf{Bunching}. Most interestingly, the perturbative Hamiltonian in Eq.~\eqref{eq:effH} implies a tendency for bosons to bunch on neighboring sites: the pair attraction is minimized for a superposition of states in which two neighboring sites are each occupied by $L/2$ bosons, $\ket{\psi_\mathrm{bunch}}=\frac{1}{\sqrt{L}}\sum_i \ket{0\cdots 0 \left(\frac{L}{2}\right)_i \left(\frac{L}{2}\right)_{i+1}0\cdots 0}$. Strikingly, this state, reminiscent of the W-like state that emerges in the attractive Bose-Hubbard-model~\cite{jack2005,Mansikkamaki2022},  has superextensive energy $\braket{H}\propto -L^2$. In a grand-canonical setting, a state with superextensive energy is not a stable phase of matter because the energy cost can never counterbalance the chemical potential cost and therefore, infinitely many particles are pulled out of the reservoir~\cite{Dyson1967} (see End Matter). Because of the almost equal strength of the pair hopping, $\ket{\psi_\mathrm{bunch}}$ is modified by the resonance between those ``maximally bunched'' states and states in which two bosons have hopped to another site.

For non-zero on-site interaction strength $U$, the highly occupied bunched states are penalized. Therefore, for sufficiently large $U$, the state should qualitatively change when the on-site interaction repulsion energy exceeds the pair hopping and attraction energy. For large $L$, we estimate this ``transition point'' $U_\mathrm{c}$ by solving $\braket{\psi_\mathrm{bunch}|\hat H|\psi_\mathrm{bunch}}=0$ for $U$, finding $U_\mathrm{c}=6 J_\mathrm{pair}$. 

To test this intuitive prediction, we make use of the reduction of the bosonic Hilbert space dimension in the unit-filling sector to $\binom{L+L/2-1}{L/2}$. Specifically, we construct the effective Hamiltonian of the bosons within the constrained Hilbert space, enabling us to simulate larger systems within exact diagonalization than in the original full LGT. DMRG struggles in this regime due to the anticipated large number fluctuations, requiring a large on-site Hilbert-space dimension.

We show the energy density $E/L$ of the ground state in Fig.~\ref{fig:clump}a for different system sizes. For large $U$, we find all system sizes to converge quickly, as expected from a typical many-body state with extensive energy. By contrast, for small $U$, we find the energy density to increase linearly with system size, in agreement with the expectation from the bunching state. The transition from extensive to super-extensive energy occurs at $U/J_\mathrm{pair}\approx 6$ as predicted from our estimate.

To confirm the bunching nature of the many-body state in this regime, we plot the system-size-normalized on-site fluctuations $(\langle \hat n_{i}^2\rangle - \langle \hat n_{i}\rangle^2)/L$ in Fig.~\ref{fig:clump}b, introduced in Ref.~\cite{Deng2008} as a bunching order parameter in the attractive Hubbard model. We find that these normalized fluctuations are approximately independent of $L$, indicating a linear growth of fluctuations with system size. By contrast, for large $U/J_\mathrm{pair}$, the fluctuations are independent of system size and therefore the normalized fluctuations decrease to zero as the system size increases. This corroborates that the state for small on-site interactions is indeed a bunching state and that the system-size-normalized fluctuations are its order parameter. In addition, we show the pair correlation function
\begin{equation}
    C_{ij}^{(2)}=\langle (\hat a^\dagger_{i})^2(\hat a_{j})^2+\mathrm{h.c.}\rangle
\end{equation}
as a function of the distance $|i-j|$ in Fig.~\ref{fig:clump}c. For small $U/J_\mathrm{pair}$, we find nearest-neighbor pair correlations which grow with system size, and beyond-nearest-neighbor correlations which rapidly decay with distance. Both observations indicate a state whose correlations are highly concentrated between close sites, in agreement with bunching. For large  $U/J_\mathrm{pair}$, we instead find correlations which are smaller for short distances, but decay more slowly, indicating a qualitative change of the state compared to small $U/J_\mathrm{pair}$. 

So far, we have studied the transition from the bunching state as $U$ is increased while the pair hopping $J_\mathrm{pair}$ is fixed. To further characterize the shape of the bunching region in the full phase diagram, we plot its ``order parameter,'' the normalized on-site fluctuations for a fixed system size in Fig.~\ref{fig:clump}d.  For large $g/J$, we find that the border of the bunching region is in agreement with the expectation $U/J=6/(g/J)$ from the perturbative calculation. For a fixed, small on-site interaction strength, we find that the fluctuations decrease rapidly around $g/J\approx 1$. This indicates that the higher-order terms neglected in the perturbative expansion in Eq.~\eqref{eq:effH} regulate the bunching behavior. In total, we therefore find a ``dome-shaped'' bunching region. The next obvious question is about the state of the system when we transition out of the bunching region for large $U/J_\mathrm{pair}$.

\begin{figure}[t]
    \centering
    \includegraphics{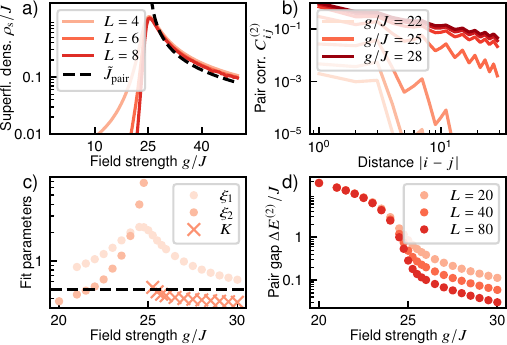}
    \caption{\textbf{Pair-superfluid-to-Mott-insulator transition.} $U/J=50$. a) Superfluid density obtained from exact diagonalization with twisted boundary conditions. b) Two-particle correlator $C^{(2)}$ from DMRG, converged with respect to system size (see End Matter). $i=L/2$, $j=L/2+1$ for $|i-j|=1$, as $|i-j|$ is increased by decreasing (increasing) $i$ ($j$) in an alternating way. c) Fit results of $C^{(1)}$ and $C^{(2)}$ to an exponential $\propto\exp(-(|i-j|/\xi_i))$ and powerlaw $\propto |i-j|^{-1/(2K)}$. d) Two particle gap calculated from DMRG.}
    \label{fig:SFMI}
\end{figure}

\textbf{Pair superfluid.}  We first consider the regime $g/J \ll U/J$. In this case, perturbative pair hopping is still favored, but with an amplitude which is reduced due to the higher energy cost of the intermediate state induced by $U$. We can estimate this amplitude by considering the states $\ket{1\downarrow 1}$ and $\ket{2\uparrow 0}$. From their energy difference $2g-U$ and transition matrix element $|\braket{2\uparrow 0|\hat H |1\downarrow 1}|^2=2J^2$, we define a modified effective pair hopping strength $\tilde J_\mathrm{pair}=\frac{2J^2}{2g-U}=\frac{J^2}{g-U/2}$. 

From this simple model, we expect that the state of the pairs is that of weakly-repulsive hopping particles, i.e. a superfluid. To test this prediction, we perform exact diagonalisation simulations and test whether the superfluid density converges to a finite value for increasing system sizes. We evaluate the superfluid density $\rho_\mathrm{s}$ in exact diagonalization by using twisted boundary conditions with a total phase $\phi$~\cite{Kiely2022}, in which case $\rho_\mathrm{s} = \frac{1}{N}\frac{\partial^2 E}{\partial\phi^2}\big|_{\phi=0}$,
where $N=L$ is the total number of bosons.

We show the superfluid density in Fig.~\ref{fig:SFMI}a. We find that for large $g/J$, it converges to a finite value as the system size increases. In addition, we find a good agreement with the prefactor $\tilde J_\mathrm{pair}$ of the effective free hopping model derived above. This shows that the system indeed enters a superfluid phase in this regime.

To show that the superfluid charge carriers are pairs of particles, we plot in Fig.~\ref{fig:SFMI}b the decay with distance of the pair correlator $C^{(2)}_{ij}$ obtained from large-scale DMRG simulations. We find an approximately power-law decay of the correlations with distance for large $g/J$. This indicates the formation of a Luttinger liquid, the hallmark 1D superfluid state. The exponent of the power-law decay varies with $g/J$ and approaches $K=1/2$ as $g\rightarrow 2U$. This is the exponent expected from a superfluid at $1/2$ filling~\cite{Giamarchi2003} - a result in agreement with our expectation that the superfluid carriers are pairs, which are half-filled at unit filling of bosons. By contrast, the single-particle correlator $C^{(1)}_{ij}$ decays exponentially, with a coherence length that increases as $g\rightarrow 2U$, but does not diverge.

\textbf{Mott insulator.} As clearly visible from both the superfluid density and the two-particle correlator, the state of the system changes dramatically around $g\approx 2U$. We find that the superfluid density vanishes as $L\rightarrow\infty$ and that the two-particle correlator decreases exponentially with distance, with a correlation length that diverges as $g\rightarrow 2U$ from below. These are indications that the system enters an insulating state. One question is whether this insulating state is a Mott insulator, i.e. an incompressible state. In the canonical ensemble, a vanishing compressibility results in a finite energy cost to add a single particle to the system. In the case of the LGT under study, a single particle addition always results in an energy cost $g/J$ when enforcing the same Gauss' law sector~\cite{Kebric2021}. Therefore, the energy cost $\Delta E^{(2)}=E_{N+1}-2E_N+E_{N-1}$ of adding a pair of particles is more interesting, where $E_{N+1}$ is the ground state for $N+1$ particles on $L$ sites. We show $\Delta E^{(2)}/J$ obtained from DMRG in Fig.~\ref{fig:SFMI}d. We find that in the pair superfluid regime, this gap vanishes as the system size is increased. By contrast, for $g<2U$ the gap converges to a finite value. This confirms the presence of a pair Mott insulator for large $U$.

The pair-superfluid-to-Mott-insulator transition is intuitively understood in the limit of $J=0$. For $U/g\rightarrow\infty$, the ground state of the bosons is a simple product state with one boson occupying every site. The gauge constraint in Eq.~\eqref{eq:Gauss} then forces the fields into a Néel state, such that the overall ground state is given by $\ket{\uparrow 1 \downarrow 1 \uparrow 1 \cdots}$. On the other hand, for small $U/g\rightarrow 0$, the gauge fields are instead in a ferromagnetic product state. The gauge constraint enforces that only even-parity boson states are allowed, all of which are degenerate in this regime. An infinitesimal $J/g$ then induces superfluidity. The transition occurs when the energy of these states becomes equal, which happens at $g=U/2$. This transition is therefore induced by the effective interactions between bosons and gauge fields induced by Gauss' law.

\begin{figure}[t]
    \centering    
    \includegraphics{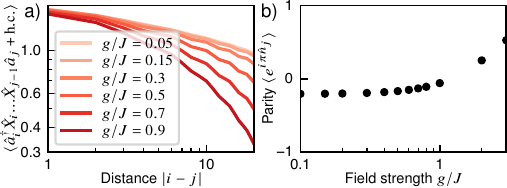}        
    \caption{\textbf{Plasma region.} DMRG results for $L=40$, $U=2.8J$. a) String-order correlator $C^{(1)}$. b) Parity in the middle of the system, $j=L/2$. }
    \label{fig_Z2_MPS}
\end{figure}

\textbf{Plasma region.} Finally, we study the regime in which the hopping dominates over field strength and on-site interaction. At first, one may guess that a single-particle superfluid is possible in this regime. However, we have already established that the single-particle correlator always decays exponentially, which excludes a true long-range correlated state. However, we numerically find that $C^{(1)}$ shows almost no decay for distances as large as $20$ sites for $g/J\lesssim 1$ at $U/J=2.8$, c.f. Fig.~\ref{fig_Z2_MPS}a, which is below the critical value $U_\mathrm{c}/J$ of the single-particle superfluid transition for $g=0$.  This indicates that single particles are mobile over short distances and that states that have single occupancies on some sites make up a large fraction of the ground state. This latter fact can be more clearly seen from the parity $\braket{e^{i\pi \hat n_j}}$ shown in Fig.~\ref{fig_Z2_MPS}b. A parity smaller than unity indicates the presence of states with odd occupation number; in particular, a lattice condensate $e^{- (1/L) \sum_j e^{-ikj} \hat a^\dagger_j} \ket{\mathrm{vacuum}}$ with (on average) unit filling has parity $\braket{e^{i\pi \hat n_j}}=e^{-2}\approx 0.14$ (see End Matter). This value is modified to lower values for finite $U$ as double occupancies are suppressed by the on-site interactions, but stays close to zero. Indeed, for $g/J\lesssim 1$, we find that the parity is close to zero and agrees well with the value $\braket{e^{i\pi \hat n_j}}\approx -0.2$ found for the Bose-Hubbard model (i.e. $g=0$). This indicates that states with even and odd parity contribute almost equally to relatively large values of $g/J$. At $g/J \approx 1$ the parity rapidly increases, indicating the crossover to the pair superfluid. Inspired by the hard-core boson case studied in Ref.~\cite{Kebric2021}, where single particles coexist with nearest-neighbor mesons, we call this phase a plasma due to its strong similarities to an electromagnetic plasma, where ions and electrons are deconfined over short distances.

\textbf{Discussion \& Outlook.} While we have established many qualitative features of the phase diagram of the $\mathbb{Z}_2$-Higgs, the most pressing open questions relate to the bunching region: Does it prevail at large system sizes and, if not, what is the ground state at large $L$? The superextensivity of its energy most likely does not prevail in large systems: higher-order terms in the perturbative expansion most likely regulate the divergence due to their stronger scaling with $N\propto L$. For small $J_\mathrm{pair}/U$, the value of $L$ where this finite-size crossover occurs can, however, be large such that the bunching can prevail to experimentally relevant scales. Another question is the relation of the effective pair physics in our model to the effective dipole physics emerging in the tilted-field Bose-Hubbard model~\cite{Sachdev2002,guardadosanchez2020,Zechmann2023} and whether constrained dynamics observed there~\cite{sala2020,Burchards2022} could emerge in the $\mathbb{Z}_2$-Higgs model, too.  Numerically, a more in-depth study of the bunching phase using DMRG simulations is, however, challenging because of its macroscopic local particle number fluctuations, thus requiring large local Hilbert space dimensions. Therefore, quantum simulators using native bosonic degrees of freedom such as trapped ions~\cite{Whitlow2023}, circuit QED~\cite{Wang2020}, and cold atoms in optical lattices~\cite{Schweizer2019} might be ideally suited to answer these questions. To this end, we showed in Ref.~\cite{crane2024} how a simulation of the $\mathbb{Z}_2$-Higgs model can be realized in circuit QED, making use of digital qubit-boson gates~\cite{liu2024} to, for example, measure the superfluid density. Finally, direct extensions of the present work relate to the finite-temperature phase diagram~\cite{kebric2023a,mueller2024,schuckert2025}, and the coupling of Higgs fields to fermionic matter, which could be studied in qubit~\cite{Hemery2024,nigmatullin2024,evered2025} or fermion-qubit quantum computers~\cite{gonzalezcuadra2023,zache2023,rad2024,schuckert2024}.
\begin{acknowledgments}
\textbf{Acknowledgements.} We acknowledge discussions with Zohreh Davoudi, Alexey Gorshkov, Fabian Grusdt, Matjaž Kebrič, and Niklas M\"uller. This material is based upon work supported by the U.S. Department of Energy, Office of Science, National Quantum Information Science Research Centers, Quantum Systems Accelerator (QSA) and Co-Design Center for Quantum Advantage (C${}^{2}$QA). QSA and C${}^{2}$QA collaborated in this research. SK is supported with funds from the Ministry of Science, Research and Culture of the State of Brandenburg within the Center for Quantum Technology and Applications (CQTA). EC~is supported in part by ARO MURI (award No.~SCON-00005095), and DoE (BNL contract No.~433702).
\end{acknowledgments}
\bibliography{bib.bib}

\begin{thebibliography}{56}%
\makeatletter
\providecommand \@ifxundefined [1]{%
 \@ifx{#1\undefined}
}%
\providecommand \@ifnum [1]{%
 \ifnum #1\expandafter \@firstoftwo
 \else \expandafter \@secondoftwo
 \fi
}%
\providecommand \@ifx [1]{%
 \ifx #1\expandafter \@firstoftwo
 \else \expandafter \@secondoftwo
 \fi
}%
\providecommand \natexlab [1]{#1}%
\providecommand \enquote  [1]{``#1''}%
\providecommand \bibnamefont  [1]{#1}%
\providecommand \bibfnamefont [1]{#1}%
\providecommand \citenamefont [1]{#1}%
\providecommand \href@noop [0]{\@secondoftwo}%
\providecommand \href [0]{\begingroup \@sanitize@url \@href}%
\providecommand \@href[1]{\@@startlink{#1}\@@href}%
\providecommand \@@href[1]{\endgroup#1\@@endlink}%
\providecommand \@sanitize@url [0]{\catcode `\\12\catcode `\$12\catcode `\&12\catcode `\#12\catcode `\^12\catcode `\_12\catcode `\%12\relax}%
\providecommand \@@startlink[1]{}%
\providecommand \@@endlink[0]{}%
\providecommand \url  [0]{\begingroup\@sanitize@url \@url }%
\providecommand \@url [1]{\endgroup\@href {#1}{\urlprefix }}%
\providecommand \urlprefix  [0]{URL }%
\providecommand \Eprint [0]{\href }%
\providecommand \doibase [0]{https://doi.org/}%
\providecommand \selectlanguage [0]{\@gobble}%
\providecommand \bibinfo  [0]{\@secondoftwo}%
\providecommand \bibfield  [0]{\@secondoftwo}%
\providecommand \translation [1]{[#1]}%
\providecommand \BibitemOpen [0]{}%
\providecommand \bibitemStop [0]{}%
\providecommand \bibitemNoStop [0]{.\EOS\space}%
\providecommand \EOS [0]{\spacefactor3000\relax}%
\providecommand \BibitemShut  [1]{\csname bibitem#1\endcsname}%
\let\auto@bib@innerbib\@empty
%</preamble>
\bibitem [{\citenamefont {Wilson}(1974)}]{Wilson1974}%
  \BibitemOpen
  \bibfield  {author} {\bibinfo {author} {\bibfnamefont {K.~G.}\ \bibnamefont {Wilson}},\ }\bibfield  {title} {\emph {\bibinfo {title} {Confinement of quarks}},\ }\href {https://doi.org/10.1103/PhysRevD.10.2445} {\bibfield  {journal} {\bibinfo  {journal} {Phys. Rev. D}\ }\textbf {\bibinfo {volume} {10}},\ \bibinfo {pages} {2445} (\bibinfo {year} {1974})}\BibitemShut {NoStop}%
\bibitem [{\citenamefont {Kogut}\ and\ \citenamefont {Susskind}(1975)}]{Kogut1975}%
  \BibitemOpen
  \bibfield  {author} {\bibinfo {author} {\bibfnamefont {J.}~\bibnamefont {Kogut}}\ and\ \bibinfo {author} {\bibfnamefont {L.}~\bibnamefont {Susskind}},\ }\bibfield  {title} {\emph {\bibinfo {title} {Hamiltonian formulation of Wilson's lattice gauge theories}},\ }\href {https://doi.org/10.1103/PhysRevD.11.395} {\bibfield  {journal} {\bibinfo  {journal} {Phys. Rev. D}\ }\textbf {\bibinfo {volume} {11}},\ \bibinfo {pages} {395} (\bibinfo {year} {1975})}\BibitemShut {NoStop}%
\bibitem [{\citenamefont {Wegner}(1971)}]{Wegner1971}%
  \BibitemOpen
  \bibfield  {author} {\bibinfo {author} {\bibfnamefont {F.~J.}\ \bibnamefont {Wegner}},\ }\bibfield  {title} {\emph {\bibinfo {title} {Duality in Generalized Ising Models and Phase Transitions without Local Order Parameters}},\ }\bibfield  {booktitle} {\emph {\bibinfo {booktitle} {J. Math. Phys.}},\ }\href {https://doi.org/10.1063/1.1665530} {\bibfield  {journal} {\bibinfo  {journal} {J. Math. Phys.}\ }\textbf {\bibinfo {volume} {12}},\ \bibinfo {pages} {2259} (\bibinfo {year} {1971})}\BibitemShut {NoStop}%
\bibitem [{\citenamefont {Horn}\ \emph {et~al.}(1979)\citenamefont {Horn}, \citenamefont {Weinstein},\ and\ \citenamefont {Yankielowicz}}]{Horn1979}%
  \BibitemOpen
  \bibfield  {author} {\bibinfo {author} {\bibfnamefont {D.}~\bibnamefont {Horn}}, \bibinfo {author} {\bibfnamefont {M.}~\bibnamefont {Weinstein}},\ and\ \bibinfo {author} {\bibfnamefont {S.}~\bibnamefont {Yankielowicz}},\ }\bibfield  {title} {\emph {\bibinfo {title} {Hamiltonian approach to {$Z(N)$} lattice gauge theories}},\ }\href {https://doi.org/10.1103/PhysRevD.19.3715} {\bibfield  {journal} {\bibinfo  {journal} {Phys. Rev. D}\ }\textbf {\bibinfo {volume} {19}},\ \bibinfo {pages} {3715} (\bibinfo {year} {1979})}\BibitemShut {NoStop}%
\bibitem [{\citenamefont {Borla}\ \emph {et~al.}(2020)\citenamefont {Borla}, \citenamefont {Verresen}, \citenamefont {Grusdt},\ and\ \citenamefont {Moroz}}]{Borla2020}%
  \BibitemOpen
  \bibfield  {author} {\bibinfo {author} {\bibfnamefont {U.}~\bibnamefont {Borla}}, \bibinfo {author} {\bibfnamefont {R.}~\bibnamefont {Verresen}}, \bibinfo {author} {\bibfnamefont {F.}~\bibnamefont {Grusdt}},\ and\ \bibinfo {author} {\bibfnamefont {S.}~\bibnamefont {Moroz}},\ }\bibfield  {title} {\emph {\bibinfo {title} {Confined Phases of One-Dimensional Spinless Fermions Coupled to ${Z}_{2}$ Gauge Theory}},\ }\href {https://doi.org/10.1103/PhysRevLett.124.120503} {\bibfield  {journal} {\bibinfo  {journal} {Phys. Rev. Lett.}\ }\textbf {\bibinfo {volume} {124}},\ \bibinfo {pages} {120503} (\bibinfo {year} {2020})}\BibitemShut {NoStop}%
\bibitem [{\citenamefont {Schweizer}\ \emph {et~al.}(2019)\citenamefont {Schweizer}, \citenamefont {Grusdt}, \citenamefont {Berngruber}, \citenamefont {Barbiero}, \citenamefont {Demler}, \citenamefont {Goldman}, \citenamefont {Bloch},\ and\ \citenamefont {Aidelsburger}}]{Schweizer2019}%
  \BibitemOpen
  \bibfield  {author} {\bibinfo {author} {\bibfnamefont {C.}~\bibnamefont {Schweizer}}, \bibinfo {author} {\bibfnamefont {F.}~\bibnamefont {Grusdt}}, \bibinfo {author} {\bibfnamefont {M.}~\bibnamefont {Berngruber}}, \bibinfo {author} {\bibfnamefont {L.}~\bibnamefont {Barbiero}}, \bibinfo {author} {\bibfnamefont {E.}~\bibnamefont {Demler}}, \bibinfo {author} {\bibfnamefont {N.}~\bibnamefont {Goldman}}, \bibinfo {author} {\bibfnamefont {I.}~\bibnamefont {Bloch}},\ and\ \bibinfo {author} {\bibfnamefont {M.}~\bibnamefont {Aidelsburger}},\ }\bibfield  {title} {\emph {\bibinfo {title} {Floquet approach to $\mathbb{Z}_2$ lattice gauge theories with ultracold atoms in optical lattices}},\ }\href {https://doi.org/10.1038/s41567-019-0649-7} {\bibfield  {journal} {\bibinfo  {journal} {Nat. Phys.}\ }\textbf {\bibinfo {volume} {15}},\ \bibinfo {pages} {1168} (\bibinfo {year} {2019})}\BibitemShut {NoStop}%
\bibitem [{\citenamefont {Barbiero}\ \emph {et~al.}(2019)\citenamefont {Barbiero}, \citenamefont {Schweizer}, \citenamefont {Aidelsburger}, \citenamefont {Demler}, \citenamefont {Goldman},\ and\ \citenamefont {Grusdt}}]{Barbiero2018}%
  \BibitemOpen
  \bibfield  {author} {\bibinfo {author} {\bibfnamefont {L.}~\bibnamefont {Barbiero}}, \bibinfo {author} {\bibfnamefont {C.}~\bibnamefont {Schweizer}}, \bibinfo {author} {\bibfnamefont {M.}~\bibnamefont {Aidelsburger}}, \bibinfo {author} {\bibfnamefont {E.}~\bibnamefont {Demler}}, \bibinfo {author} {\bibfnamefont {N.}~\bibnamefont {Goldman}},\ and\ \bibinfo {author} {\bibfnamefont {F.}~\bibnamefont {Grusdt}},\ }\bibfield  {title} {\emph {\bibinfo {title} {Coupling ultracold matter to dynamical gauge fields in optical lattices: From flux attachment to $\mathbb{Z}_2$ lattice gauge theories}},\ }\href {http://dx.doi.org/10.1126/sciadv.aav7444} {\bibfield  {journal} {\bibinfo  {journal} {Sci. Adv.}\ }\textbf {\bibinfo {volume} {5}} (\bibinfo {year} {2019})}\BibitemShut {NoStop}%
\bibitem [{\citenamefont {Bǎzǎvan}\ \emph {et~al.}(2024)\citenamefont {Bǎzǎvan}, \citenamefont {Saner}, \citenamefont {Tirrito}, \citenamefont {Araneda}, \citenamefont {Srinivas},\ and\ \citenamefont {Bermudez}}]{bazavan2023}%
  \BibitemOpen
  \bibfield  {author} {\bibinfo {author} {\bibfnamefont {O.}~\bibnamefont {Bǎzǎvan}}, \bibinfo {author} {\bibfnamefont {S.}~\bibnamefont {Saner}}, \bibinfo {author} {\bibfnamefont {E.}~\bibnamefont {Tirrito}}, \bibinfo {author} {\bibfnamefont {G.}~\bibnamefont {Araneda}}, \bibinfo {author} {\bibfnamefont {R.}~\bibnamefont {Srinivas}},\ and\ \bibinfo {author} {\bibfnamefont {A.}~\bibnamefont {Bermudez}},\ }\bibfield  {title} {\emph {\bibinfo {title} {Synthetic $Z_2$ gauge theories based on parametric excitations of trapped ions}},\ }\href {https://doi.org/10.1038/s42005-024-01691-w} {\bibfield  {journal} {\bibinfo  {journal} {Comm. Phys.}\ }\textbf {\bibinfo {volume} {7}},\ \bibinfo {pages} {229} (\bibinfo {year} {2024})}\BibitemShut {NoStop}%
\bibitem [{\citenamefont {Davoudi}\ \emph {et~al.}(2023)\citenamefont {Davoudi}, \citenamefont {Mueller},\ and\ \citenamefont {Powers}}]{Davoudi2023}%
  \BibitemOpen
  \bibfield  {author} {\bibinfo {author} {\bibfnamefont {Z.}~\bibnamefont {Davoudi}}, \bibinfo {author} {\bibfnamefont {N.}~\bibnamefont {Mueller}},\ and\ \bibinfo {author} {\bibfnamefont {C.}~\bibnamefont {Powers}},\ }\bibfield  {title} {\emph {\bibinfo {title} {Towards Quantum Computing Phase Diagrams of Gauge Theories with Thermal Pure Quantum States}},\ }\href {https://doi.org/10.1103/PhysRevLett.131.081901} {\bibfield  {journal} {\bibinfo  {journal} {Phys. Rev. Lett.}\ }\textbf {\bibinfo {volume} {131}},\ \bibinfo {pages} {081901} (\bibinfo {year} {2023})}\BibitemShut {NoStop}%
\bibitem [{\citenamefont {Bennewitz}\ \emph {et~al.}(2024)\citenamefont {Bennewitz}, \citenamefont {Ware}, \citenamefont {Schuckert}, \citenamefont {Lerose}, \citenamefont {Surace}, \citenamefont {Belyansky}, \citenamefont {Morong}, \citenamefont {Luo}, \citenamefont {De}, \citenamefont {Collins}, \citenamefont {Katz}, \citenamefont {Monroe}, \citenamefont {Davoudi},\ and\ \citenamefont {Gorshkov}}]{bennewitz2024}%
  \BibitemOpen
  \bibfield  {author} {\bibinfo {author} {\bibfnamefont {E.~R.}\ \bibnamefont {Bennewitz}}, \bibinfo {author} {\bibfnamefont {B.}~\bibnamefont {Ware}}, \bibinfo {author} {\bibfnamefont {A.}~\bibnamefont {Schuckert}}, \bibinfo {author} {\bibfnamefont {A.}~\bibnamefont {Lerose}}, \bibinfo {author} {\bibfnamefont {F.~M.}\ \bibnamefont {Surace}}, \bibinfo {author} {\bibfnamefont {R.}~\bibnamefont {Belyansky}}, \bibinfo {author} {\bibfnamefont {W.}~\bibnamefont {Morong}}, \bibinfo {author} {\bibfnamefont {D.}~\bibnamefont {Luo}}, \bibinfo {author} {\bibfnamefont {A.}~\bibnamefont {De}}, \bibinfo {author} {\bibfnamefont {K.~S.}\ \bibnamefont {Collins}}, \bibinfo {author} {\bibfnamefont {O.}~\bibnamefont {Katz}}, \bibinfo {author} {\bibfnamefont {C.}~\bibnamefont {Monroe}}, \bibinfo {author} {\bibfnamefont {Z.}~\bibnamefont {Davoudi}},\ and\ \bibinfo {author} {\bibfnamefont {A.~V.}\ \bibnamefont {Gorshkov}},\ }\href@noop {} {\bibfield  {title} {\emph {\bibinfo {title} {Simulating Meson Scattering on Spin Quantum
  Simulators}}}} (\bibinfo {year} {2024}),\ \Eprint {https://arxiv.org/abs/2403.07061} {arXiv:2403.07061} \BibitemShut {NoStop}%
\bibitem [{\citenamefont {De}\ \emph {et~al.}(2024)\citenamefont {De}, \citenamefont {Lerose}, \citenamefont {Luo}, \citenamefont {Surace}, \citenamefont {Schuckert}, \citenamefont {Bennewitz}, \citenamefont {Ware}, \citenamefont {Morong}, \citenamefont {Collins}, \citenamefont {Davoudi}, \citenamefont {Gorshkov}, \citenamefont {Katz},\ and\ \citenamefont {Monroe}}]{de2024}%
  \BibitemOpen
  \bibfield  {author} {\bibinfo {author} {\bibfnamefont {A.}~\bibnamefont {De}}, \bibinfo {author} {\bibfnamefont {A.}~\bibnamefont {Lerose}}, \bibinfo {author} {\bibfnamefont {D.}~\bibnamefont {Luo}}, \bibinfo {author} {\bibfnamefont {F.~M.}\ \bibnamefont {Surace}}, \bibinfo {author} {\bibfnamefont {A.}~\bibnamefont {Schuckert}}, \bibinfo {author} {\bibfnamefont {E.~R.}\ \bibnamefont {Bennewitz}}, \bibinfo {author} {\bibfnamefont {B.}~\bibnamefont {Ware}}, \bibinfo {author} {\bibfnamefont {W.}~\bibnamefont {Morong}}, \bibinfo {author} {\bibfnamefont {K.~S.}\ \bibnamefont {Collins}}, \bibinfo {author} {\bibfnamefont {Z.}~\bibnamefont {Davoudi}}, \bibinfo {author} {\bibfnamefont {A.~V.}\ \bibnamefont {Gorshkov}}, \bibinfo {author} {\bibfnamefont {O.}~\bibnamefont {Katz}},\ and\ \bibinfo {author} {\bibfnamefont {C.}~\bibnamefont {Monroe}},\ }\href@noop {} {\bibfield  {title} {\emph {\bibinfo {title} {Observation of string-breaking dynamics in a quantum simulator}}}} (\bibinfo {year} {2024}),\ \Eprint
  {https://arxiv.org/abs/2410.13815} {arXiv:2410.13815} \BibitemShut {NoStop}%
\bibitem [{\citenamefont {Surace}\ \emph {et~al.}(2024)\citenamefont {Surace}, \citenamefont {Lerose}, \citenamefont {Katz}, \citenamefont {Bennewitz}, \citenamefont {Schuckert}, \citenamefont {Luo}, \citenamefont {De}, \citenamefont {Ware}, \citenamefont {Morong}, \citenamefont {Collins}, \citenamefont {Monroe}, \citenamefont {Davoudi},\ and\ \citenamefont {Gorshkov}}]{surace2024}%
  \BibitemOpen
  \bibfield  {author} {\bibinfo {author} {\bibfnamefont {F.~M.}\ \bibnamefont {Surace}}, \bibinfo {author} {\bibfnamefont {A.}~\bibnamefont {Lerose}}, \bibinfo {author} {\bibfnamefont {O.}~\bibnamefont {Katz}}, \bibinfo {author} {\bibfnamefont {E.~R.}\ \bibnamefont {Bennewitz}}, \bibinfo {author} {\bibfnamefont {A.}~\bibnamefont {Schuckert}}, \bibinfo {author} {\bibfnamefont {D.}~\bibnamefont {Luo}}, \bibinfo {author} {\bibfnamefont {A.}~\bibnamefont {De}}, \bibinfo {author} {\bibfnamefont {B.}~\bibnamefont {Ware}}, \bibinfo {author} {\bibfnamefont {W.}~\bibnamefont {Morong}}, \bibinfo {author} {\bibfnamefont {K.}~\bibnamefont {Collins}}, \bibinfo {author} {\bibfnamefont {C.}~\bibnamefont {Monroe}}, \bibinfo {author} {\bibfnamefont {Z.}~\bibnamefont {Davoudi}},\ and\ \bibinfo {author} {\bibfnamefont {A.~V.}\ \bibnamefont {Gorshkov}},\ }\href {https://arxiv.org/abs/2411.10652} {\bibfield  {title} {\emph {\bibinfo {title} {String-Breaking Dynamics in Quantum Adiabatic and Diabatic Processes}}}} (\bibinfo {year}
  {2024}),\ \Eprint {https://arxiv.org/abs/2411.10652} {arXiv:2411.10652 [quant-ph]} \BibitemShut {NoStop}%
\bibitem [{\citenamefont {Mildenberger}\ \emph {et~al.}(2025)\citenamefont {Mildenberger}, \citenamefont {Mruczkiewicz}, \citenamefont {Halimeh}, \citenamefont {Jiang},\ and\ \citenamefont {Hauke}}]{mildenberger2025}%
  \BibitemOpen
  \bibfield  {author} {\bibinfo {author} {\bibfnamefont {J.}~\bibnamefont {Mildenberger}}, \bibinfo {author} {\bibfnamefont {W.}~\bibnamefont {Mruczkiewicz}}, \bibinfo {author} {\bibfnamefont {J.~C.}\ \bibnamefont {Halimeh}}, \bibinfo {author} {\bibfnamefont {Z.}~\bibnamefont {Jiang}},\ and\ \bibinfo {author} {\bibfnamefont {P.}~\bibnamefont {Hauke}},\ }\bibfield  {title} {\emph {\bibinfo {title} {Confinement in a $\mathbb{Z}_2$ Lattice Gauge Theory on a Quantum Computer}},\ }\href {https://doi.org/10.1038/s41567-024-02723-6} {\bibfield  {journal} {\bibinfo  {journal} {Nat. Phys.}\ } (\bibinfo {year} {2025})}\BibitemShut {NoStop}%
\bibitem [{\citenamefont {Mueller}\ \emph {et~al.}(2024)\citenamefont {Mueller}, \citenamefont {Wang}, \citenamefont {Katz}, \citenamefont {Davoudi},\ and\ \citenamefont {Cetina}}]{mueller2024}%
  \BibitemOpen
  \bibfield  {author} {\bibinfo {author} {\bibfnamefont {N.}~\bibnamefont {Mueller}}, \bibinfo {author} {\bibfnamefont {T.}~\bibnamefont {Wang}}, \bibinfo {author} {\bibfnamefont {O.}~\bibnamefont {Katz}}, \bibinfo {author} {\bibfnamefont {Z.}~\bibnamefont {Davoudi}},\ and\ \bibinfo {author} {\bibfnamefont {M.}~\bibnamefont {Cetina}},\ }\href@noop {} {\bibfield  {title} {\emph {\bibinfo {title} {Quantum Computing Universal Thermalization Dynamics in a (2+1)D Lattice Gauge Theory}}}} (\bibinfo {year} {2024}),\ \Eprint {https://arxiv.org/abs/2408.00069} {arXiv:2408.00069} \BibitemShut {NoStop}%
\bibitem [{\citenamefont {Cochran}\ \emph {et~al.}(2024)\citenamefont {Cochran}, \citenamefont {Jobst}, \citenamefont {Rosenberg}, \citenamefont {Lensky}, \citenamefont {Gyawali}, \citenamefont {Eassa}, \citenamefont {Will}, \citenamefont {Abanin}, \citenamefont {Acharya}, \citenamefont {Beni}, \citenamefont {Andersen}, \citenamefont {Ansmann}, \citenamefont {Arute}, \citenamefont {Arya}, \citenamefont {Asfaw}, \citenamefont {Atalaya}, \citenamefont {Babbush}, \citenamefont {Ballard}, \citenamefont {Bardin}, \citenamefont {Bengtsson}, \citenamefont {Bilmes}, \citenamefont {Bourassa}, \citenamefont {Bovaird}, \citenamefont {Broughton}, \citenamefont {Browne}, \citenamefont {Buchea}, \citenamefont {Buckley}, \citenamefont {Burger}, \citenamefont {Burkett}, \citenamefont {Bushnell}, \citenamefont {Cabrera}, \citenamefont {Campero}, \citenamefont {Chang}, \citenamefont {Chen}, \citenamefont {Chiaro}, \citenamefont {Claes}, \citenamefont {Cleland}, \citenamefont {Cogan}, \citenamefont {Collins}, \citenamefont
  {Conner}, \citenamefont {Courtney}, \citenamefont {Crook}, \citenamefont {Curtin}, \citenamefont {Das}, \citenamefont {Demura}, \citenamefont {Lorenzo}, \citenamefont {Paolo}, \citenamefont {Donohoe}, \citenamefont {Drozdov}, \citenamefont {Dunsworth}, \citenamefont {Eickbusch}, \citenamefont {Elbag}, \citenamefont {Elzouka}, \citenamefont {Erickson}, \citenamefont {Ferreira}, \citenamefont {Burgos}, \citenamefont {Forati}, \citenamefont {Fowler}, \citenamefont {Foxen}, \citenamefont {Ganjam}, \citenamefont {Gasca}, \citenamefont {Élie Genois}, \citenamefont {Giang}, \citenamefont {Gilboa}, \citenamefont {Gosula}, \citenamefont {Dau}, \citenamefont {Graumann}, \citenamefont {Greene}, \citenamefont {Gross}, \citenamefont {Habegger}, \citenamefont {Hansen}, \citenamefont {Harrigan}, \citenamefont {Harrington}, \citenamefont {Heu}, \citenamefont {Higgott}, \citenamefont {Hilton}, \citenamefont {Huang}, \citenamefont {Huff}, \citenamefont {Huggins}, \citenamefont {Jeffrey}, \citenamefont {Jiang}, \citenamefont
  {Jones}, \citenamefont {Joshi}, \citenamefont {Juhas}, \citenamefont {Kafri}, \citenamefont {Kang}, \citenamefont {Karamlou}, \citenamefont {Kechedzhi}, \citenamefont {Khaire}, \citenamefont {Khattar}, \citenamefont {Khezri}, \citenamefont {Kim}, \citenamefont {Klimov}, \citenamefont {Kobrin}, \citenamefont {Korotkov}, \citenamefont {Kostritsa}, \citenamefont {Kreikebaum}, \citenamefont {Kurilovich}, \citenamefont {Landhuis}, \citenamefont {Lange-Dei}, \citenamefont {Langley}, \citenamefont {Lau}, \citenamefont {Ledford}, \citenamefont {Lee}, \citenamefont {Lester}, \citenamefont {Guevel}, \citenamefont {Li}, \citenamefont {Lill}, \citenamefont {Livingston}, \citenamefont {Locharla}, \citenamefont {Lundahl}, \citenamefont {Lunt}, \citenamefont {Madhuk}, \citenamefont {Maloney}, \citenamefont {Mandrà}, \citenamefont {Martin}, \citenamefont {Martin}, \citenamefont {Maxfield}, \citenamefont {McClean}, \citenamefont {McEwen}, \citenamefont {Meeks}, \citenamefont {Megrant}, \citenamefont {Miao}, \citenamefont
  {Molavi}, \citenamefont {Molina}, \citenamefont {Montazeri}, \citenamefont {Movassagh}, \citenamefont {Neill}, \citenamefont {Newman}, \citenamefont {Nguyen}, \citenamefont {Nguyen}, \citenamefont {Ni}, \citenamefont {Niu}, \citenamefont {Oliver}, \citenamefont {Ottosson}, \citenamefont {Pizzuto}, \citenamefont {Potter}, \citenamefont {Pritchard}, \citenamefont {Quintana}, \citenamefont {Ramachandran}, \citenamefont {Reagor}, \citenamefont {Rhodes}, \citenamefont {Roberts}, \citenamefont {Sankaragomathi}, \citenamefont {Satzinger}, \citenamefont {Schurkus}, \citenamefont {Shearn}, \citenamefont {Shorter}, \citenamefont {Shutty}, \citenamefont {Shvarts}, \citenamefont {Sivak}, \citenamefont {Small}, \citenamefont {Smith}, \citenamefont {Springer}, \citenamefont {Sterling}, \citenamefont {Suchard}, \citenamefont {Szasz}, \citenamefont {Sztein}, \citenamefont {Thor}, \citenamefont {Torunbalci}, \citenamefont {Vaishnav}, \citenamefont {Vargas}, \citenamefont {Vdovichev}, \citenamefont {Vidal}, \citenamefont
  {Heidweiller}, \citenamefont {Waltman}, \citenamefont {Wang}, \citenamefont {Ware}, \citenamefont {White}, \citenamefont {Wong}, \citenamefont {Woo}, \citenamefont {Xing}, \citenamefont {Yao}, \citenamefont {Yeh}, \citenamefont {Ying}, \citenamefont {Yoo}, \citenamefont {Yosri}, \citenamefont {Young}, \citenamefont {Zalcman}, \citenamefont {Zhang}, \citenamefont {Zhu}, \citenamefont {Zobris}, \citenamefont {Boixo}, \citenamefont {Kelly}, \citenamefont {Lucero}, \citenamefont {Chen}, \citenamefont {Smelyanskiy}, \citenamefont {Neven}, \citenamefont {Gammon-Smith}, \citenamefont {Pollmann}, \citenamefont {Knap},\ and\ \citenamefont {Roushan}}]{cochran2024}%
  \BibitemOpen
  \bibfield  {author} {\bibinfo {author} {\bibfnamefont {T.~A.}\ \bibnamefont {Cochran}}, \bibinfo {author} {\bibfnamefont {B.}~\bibnamefont {Jobst}}, \bibinfo {author} {\bibfnamefont {E.}~\bibnamefont {Rosenberg}}, \bibinfo {author} {\bibfnamefont {Y.~D.}\ \bibnamefont {Lensky}}, \bibinfo {author} {\bibfnamefont {G.}~\bibnamefont {Gyawali}}, \bibinfo {author} {\bibfnamefont {N.}~\bibnamefont {Eassa}}, \bibinfo {author} {\bibfnamefont {M.}~\bibnamefont {Will}}, \bibinfo {author} {\bibfnamefont {D.}~\bibnamefont {Abanin}}, \bibinfo {author} {\bibfnamefont {R.}~\bibnamefont {Acharya}}, \bibinfo {author} {\bibfnamefont {L.~A.}\ \bibnamefont {Beni}}, \bibinfo {author} {\bibfnamefont {T.~I.}\ \bibnamefont {Andersen}}, \bibinfo {author} {\bibfnamefont {M.}~\bibnamefont {Ansmann}}, \bibinfo {author} {\bibfnamefont {F.}~\bibnamefont {Arute}}, \bibinfo {author} {\bibfnamefont {K.}~\bibnamefont {Arya}}, \bibinfo {author} {\bibfnamefont {A.}~\bibnamefont {Asfaw}}, \bibinfo {author} {\bibfnamefont {J.}~\bibnamefont
  {Atalaya}}, \bibinfo {author} {\bibfnamefont {R.}~\bibnamefont {Babbush}}, \bibinfo {author} {\bibfnamefont {B.}~\bibnamefont {Ballard}}, \bibinfo {author} {\bibfnamefont {J.~C.}\ \bibnamefont {Bardin}}, \emph {et~al.},\ }\href@noop {} {\bibfield  {title} {\emph {\bibinfo {title} {Visualizing Dynamics of Charges and Strings in (2+1)D Lattice Gauge Theories}}}} (\bibinfo {year} {2024}),\ \Eprint {https://arxiv.org/abs/2409.17142} {arXiv:2409.17142} \BibitemShut {NoStop}%
\bibitem [{\citenamefont {Gonzalez-Cuadra}\ \emph {et~al.}(2024)\citenamefont {Gonzalez-Cuadra}, \citenamefont {Hamdan}, \citenamefont {Zache}, \citenamefont {Braverman}, \citenamefont {Kornjaca}, \citenamefont {Lukin}, \citenamefont {Cantu}, \citenamefont {Liu}, \citenamefont {Wang}, \citenamefont {Keesling}, \citenamefont {Lukin}, \citenamefont {Zoller},\ and\ \citenamefont {Bylinskii}}]{gonzalezcuadra2024}%
  \BibitemOpen
  \bibfield  {author} {\bibinfo {author} {\bibfnamefont {D.}~\bibnamefont {Gonzalez-Cuadra}}, \bibinfo {author} {\bibfnamefont {M.}~\bibnamefont {Hamdan}}, \bibinfo {author} {\bibfnamefont {T.~V.}\ \bibnamefont {Zache}}, \bibinfo {author} {\bibfnamefont {B.}~\bibnamefont {Braverman}}, \bibinfo {author} {\bibfnamefont {M.}~\bibnamefont {Kornjaca}}, \bibinfo {author} {\bibfnamefont {A.}~\bibnamefont {Lukin}}, \bibinfo {author} {\bibfnamefont {S.~H.}\ \bibnamefont {Cantu}}, \bibinfo {author} {\bibfnamefont {F.}~\bibnamefont {Liu}}, \bibinfo {author} {\bibfnamefont {S.-T.}\ \bibnamefont {Wang}}, \bibinfo {author} {\bibfnamefont {A.}~\bibnamefont {Keesling}}, \bibinfo {author} {\bibfnamefont {M.~D.}\ \bibnamefont {Lukin}}, \bibinfo {author} {\bibfnamefont {P.}~\bibnamefont {Zoller}},\ and\ \bibinfo {author} {\bibfnamefont {A.}~\bibnamefont {Bylinskii}},\ }\href@noop {} {\bibfield  {title} {\emph {\bibinfo {title} {Observation of string breaking on a (2 + 1)D Rydberg quantum simulator}}}} (\bibinfo {year} {2024}),\
  \Eprint {https://arxiv.org/abs/2410.16558} {arXiv:2410.16558} \BibitemShut {NoStop}%
\bibitem [{\citenamefont {Borla}\ \emph {et~al.}(2025)\citenamefont {Borla}, \citenamefont {Osborne}, \citenamefont {Moroz},\ and\ \citenamefont {Halimeh}}]{borla2025}%
  \BibitemOpen
  \bibfield  {author} {\bibinfo {author} {\bibfnamefont {U.}~\bibnamefont {Borla}}, \bibinfo {author} {\bibfnamefont {J.~J.}\ \bibnamefont {Osborne}}, \bibinfo {author} {\bibfnamefont {S.}~\bibnamefont {Moroz}},\ and\ \bibinfo {author} {\bibfnamefont {J.~C.}\ \bibnamefont {Halimeh}},\ }\href@noop {} {\bibfield  {title} {\emph {\bibinfo {title} {String Breaking in a $2+1$D $\mathbb{Z}_2$ Lattice Gauge Theory}}}} (\bibinfo {year} {2025}),\ \Eprint {https://arxiv.org/abs/2501.17929} {arXiv:2501.17929} \BibitemShut {NoStop}%
\bibitem [{\citenamefont {Chanda}\ \emph {et~al.}(2020)\citenamefont {Chanda}, \citenamefont {Zakrzewski}, \citenamefont {Lewenstein},\ and\ \citenamefont {Tagliacozzo}}]{Chanda2020}%
  \BibitemOpen
  \bibfield  {author} {\bibinfo {author} {\bibfnamefont {T.}~\bibnamefont {Chanda}}, \bibinfo {author} {\bibfnamefont {J.}~\bibnamefont {Zakrzewski}}, \bibinfo {author} {\bibfnamefont {M.}~\bibnamefont {Lewenstein}},\ and\ \bibinfo {author} {\bibfnamefont {L.}~\bibnamefont {Tagliacozzo}},\ }\bibfield  {title} {\emph {\bibinfo {title} {Confinement and Lack of Thermalization after Quenches in the Bosonic Schwinger Model}},\ }\href {https://doi.org/10.1103/PhysRevLett.124.180602} {\bibfield  {journal} {\bibinfo  {journal} {Phys. Rev. Lett.}\ }\textbf {\bibinfo {volume} {124}},\ \bibinfo {pages} {180602} (\bibinfo {year} {2020})}\BibitemShut {NoStop}%
\bibitem [{\citenamefont {Chanda}\ \emph {et~al.}(2022{\natexlab{a}})\citenamefont {Chanda}, \citenamefont {Lewenstein}, \citenamefont {Zakrzewski},\ and\ \citenamefont {Tagliacozzo}}]{Chanda2022a}%
  \BibitemOpen
  \bibfield  {author} {\bibinfo {author} {\bibfnamefont {T.}~\bibnamefont {Chanda}}, \bibinfo {author} {\bibfnamefont {M.}~\bibnamefont {Lewenstein}}, \bibinfo {author} {\bibfnamefont {J.}~\bibnamefont {Zakrzewski}},\ and\ \bibinfo {author} {\bibfnamefont {L.}~\bibnamefont {Tagliacozzo}},\ }\bibfield  {title} {\emph {\bibinfo {title} {Phase Diagram of $1+1\mathrm{D}$ Abelian-Higgs Model and Its Critical Point}},\ }\href {https://doi.org/10.1103/PhysRevLett.128.090601} {\bibfield  {journal} {\bibinfo  {journal} {Phys. Rev. Lett.}\ }\textbf {\bibinfo {volume} {128}},\ \bibinfo {pages} {090601} (\bibinfo {year} {2022}{\natexlab{a}})}\BibitemShut {NoStop}%
\bibitem [{\citenamefont {Gonz\'alez-Cuadra}\ \emph {et~al.}(2019)\citenamefont {Gonz\'alez-Cuadra}, \citenamefont {Dauphin}, \citenamefont {Grzybowski}, \citenamefont {W\'ojcik}, \citenamefont {Lewenstein},\ and\ \citenamefont {Bermudez}}]{gonzalezcuadra2019}%
  \BibitemOpen
  \bibfield  {author} {\bibinfo {author} {\bibfnamefont {D.}~\bibnamefont {Gonz\'alez-Cuadra}}, \bibinfo {author} {\bibfnamefont {A.}~\bibnamefont {Dauphin}}, \bibinfo {author} {\bibfnamefont {P.~R.}\ \bibnamefont {Grzybowski}}, \bibinfo {author} {\bibfnamefont {P.}~\bibnamefont {W\'ojcik}}, \bibinfo {author} {\bibfnamefont {M.}~\bibnamefont {Lewenstein}},\ and\ \bibinfo {author} {\bibfnamefont {A.}~\bibnamefont {Bermudez}},\ }\bibfield  {title} {\emph {\bibinfo {title} {Symmetry-breaking topological insulators in the ${\mathbb{Z}}_{2}$ Bose-Hubbard model}},\ }\href {https://doi.org/10.1103/PhysRevB.99.045139} {\bibfield  {journal} {\bibinfo  {journal} {Phys. Rev. B}\ }\textbf {\bibinfo {volume} {99}},\ \bibinfo {pages} {045139} (\bibinfo {year} {2019})}\BibitemShut {NoStop}%
\bibitem [{\citenamefont {Chanda}\ \emph {et~al.}(2022{\natexlab{b}})\citenamefont {Chanda}, \citenamefont {González-Cuadra}, \citenamefont {Lewenstein}, \citenamefont {Tagliacozzo},\ and\ \citenamefont {Zakrzewski}}]{chanda2022}%
  \BibitemOpen
  \bibfield  {author} {\bibinfo {author} {\bibfnamefont {T.}~\bibnamefont {Chanda}}, \bibinfo {author} {\bibfnamefont {D.}~\bibnamefont {González-Cuadra}}, \bibinfo {author} {\bibfnamefont {M.}~\bibnamefont {Lewenstein}}, \bibinfo {author} {\bibfnamefont {L.}~\bibnamefont {Tagliacozzo}},\ and\ \bibinfo {author} {\bibfnamefont {J.}~\bibnamefont {Zakrzewski}},\ }\bibfield  {title} {\emph {\bibinfo {title} {{Devil's staircase of topological Peierls insulators and Peierls supersolids}}},\ }\href {https://doi.org/10.21468/SciPostPhys.12.2.076} {\bibfield  {journal} {\bibinfo  {journal} {SciPost Phys.}\ }\textbf {\bibinfo {volume} {12}},\ \bibinfo {pages} {076} (\bibinfo {year} {2022}{\natexlab{b}})}\BibitemShut {NoStop}%
\bibitem [{\citenamefont {Watanabe}\ \emph {et~al.}(2025)\citenamefont {Watanabe}, \citenamefont {Watabe},\ and\ \citenamefont {Nikuni}}]{watanabe2025}%
  \BibitemOpen
  \bibfield  {author} {\bibinfo {author} {\bibfnamefont {Y.}~\bibnamefont {Watanabe}}, \bibinfo {author} {\bibfnamefont {S.}~\bibnamefont {Watabe}},\ and\ \bibinfo {author} {\bibfnamefont {T.}~\bibnamefont {Nikuni}},\ }\href@noop {} {\bibfield  {title} {\emph {\bibinfo {title} {Transverse Field Dependence of the Ground State in the Z2 Bose-Hubbard Model}}}} (\bibinfo {year} {2025}),\ \Eprint {https://arxiv.org/abs/2501.15490} {arXiv:2501.15490} \BibitemShut {NoStop}%
\bibitem [{\citenamefont {Senthil}\ and\ \citenamefont {Fisher}(2000)}]{Senthil2000}%
  \BibitemOpen
  \bibfield  {author} {\bibinfo {author} {\bibfnamefont {T.}~\bibnamefont {Senthil}}\ and\ \bibinfo {author} {\bibfnamefont {M.~P.~A.}\ \bibnamefont {Fisher}},\ }\bibfield  {title} {\emph {\bibinfo {title} {${Z}_{2}$ gauge theory of electron fractionalization in strongly correlated systems}},\ }\href {https://doi.org/10.1103/PhysRevB.62.7850} {\bibfield  {journal} {\bibinfo  {journal} {Phys. Rev. B}\ }\textbf {\bibinfo {volume} {62}},\ \bibinfo {pages} {7850} (\bibinfo {year} {2000})}\BibitemShut {NoStop}%
\bibitem [{\citenamefont {Marcos}\ \emph {et~al.}(2013)\citenamefont {Marcos}, \citenamefont {Rabl}, \citenamefont {Rico},\ and\ \citenamefont {Zoller}}]{Marcos2013}%
  \BibitemOpen
  \bibfield  {author} {\bibinfo {author} {\bibfnamefont {D.}~\bibnamefont {Marcos}}, \bibinfo {author} {\bibfnamefont {P.}~\bibnamefont {Rabl}}, \bibinfo {author} {\bibfnamefont {E.}~\bibnamefont {Rico}},\ and\ \bibinfo {author} {\bibfnamefont {P.}~\bibnamefont {Zoller}},\ }\bibfield  {title} {\emph {\bibinfo {title} {Superconducting Circuits for Quantum Simulation of Dynamical Gauge Fields}},\ }\href {https://doi.org/10.1103/PhysRevLett.111.110504} {\bibfield  {journal} {\bibinfo  {journal} {Phys. Rev. Lett.}\ }\textbf {\bibinfo {volume} {111}},\ \bibinfo {pages} {110504} (\bibinfo {year} {2013})}\BibitemShut {NoStop}%
\bibitem [{\citenamefont {Davoudi}\ \emph {et~al.}(2020)\citenamefont {Davoudi}, \citenamefont {Hafezi}, \citenamefont {Monroe}, \citenamefont {Pagano}, \citenamefont {Seif},\ and\ \citenamefont {Shaw}}]{davoudi2020}%
  \BibitemOpen
  \bibfield  {author} {\bibinfo {author} {\bibfnamefont {Z.}~\bibnamefont {Davoudi}}, \bibinfo {author} {\bibfnamefont {M.}~\bibnamefont {Hafezi}}, \bibinfo {author} {\bibfnamefont {C.}~\bibnamefont {Monroe}}, \bibinfo {author} {\bibfnamefont {G.}~\bibnamefont {Pagano}}, \bibinfo {author} {\bibfnamefont {A.}~\bibnamefont {Seif}},\ and\ \bibinfo {author} {\bibfnamefont {A.}~\bibnamefont {Shaw}},\ }\bibfield  {title} {\emph {\bibinfo {title} {Towards analog quantum simulations of lattice gauge theories with trapped ions}},\ }\href {https://doi.org/10.1103/PhysRevResearch.2.023015} {\bibfield  {journal} {\bibinfo  {journal} {Phys. Rev. Res.}\ }\textbf {\bibinfo {volume} {2}},\ \bibinfo {pages} {023015} (\bibinfo {year} {2020})}\BibitemShut {NoStop}%
\bibitem [{\citenamefont {Belyansky}\ \emph {et~al.}(2024)\citenamefont {Belyansky}, \citenamefont {Whitsitt}, \citenamefont {Mueller}, \citenamefont {Fahimniya}, \citenamefont {Bennewitz}, \citenamefont {Davoudi},\ and\ \citenamefont {Gorshkov}}]{Belyansky2024}%
  \BibitemOpen
  \bibfield  {author} {\bibinfo {author} {\bibfnamefont {R.}~\bibnamefont {Belyansky}}, \bibinfo {author} {\bibfnamefont {S.}~\bibnamefont {Whitsitt}}, \bibinfo {author} {\bibfnamefont {N.}~\bibnamefont {Mueller}}, \bibinfo {author} {\bibfnamefont {A.}~\bibnamefont {Fahimniya}}, \bibinfo {author} {\bibfnamefont {E.~R.}\ \bibnamefont {Bennewitz}}, \bibinfo {author} {\bibfnamefont {Z.}~\bibnamefont {Davoudi}},\ and\ \bibinfo {author} {\bibfnamefont {A.~V.}\ \bibnamefont {Gorshkov}},\ }\bibfield  {title} {\emph {\bibinfo {title} {High-Energy Collision of Quarks and Mesons in the Schwinger Model: From Tensor Networks to Circuit QED}},\ }\href {https://doi.org/10.1103/PhysRevLett.132.091903} {\bibfield  {journal} {\bibinfo  {journal} {Phys. Rev. Lett.}\ }\textbf {\bibinfo {volume} {132}},\ \bibinfo {pages} {091903} (\bibinfo {year} {2024})}\BibitemShut {NoStop}%
\bibitem [{\citenamefont {Crane}\ \emph {et~al.}(2024)\citenamefont {Crane}, \citenamefont {Smith}, \citenamefont {Tomesh}, \citenamefont {Eickbusch}, \citenamefont {Martyn}, \citenamefont {Kühn}, \citenamefont {Funcke}, \citenamefont {DeMarco}, \citenamefont {Chuang}, \citenamefont {Wiebe}, \citenamefont {Schuckert},\ and\ \citenamefont {Girvin}}]{crane2024}%
  \BibitemOpen
  \bibfield  {author} {\bibinfo {author} {\bibfnamefont {E.}~\bibnamefont {Crane}}, \bibinfo {author} {\bibfnamefont {K.~C.}\ \bibnamefont {Smith}}, \bibinfo {author} {\bibfnamefont {T.}~\bibnamefont {Tomesh}}, \bibinfo {author} {\bibfnamefont {A.}~\bibnamefont {Eickbusch}}, \bibinfo {author} {\bibfnamefont {J.~M.}\ \bibnamefont {Martyn}}, \bibinfo {author} {\bibfnamefont {S.}~\bibnamefont {Kühn}}, \bibinfo {author} {\bibfnamefont {L.}~\bibnamefont {Funcke}}, \bibinfo {author} {\bibfnamefont {M.~A.}\ \bibnamefont {DeMarco}}, \bibinfo {author} {\bibfnamefont {I.~L.}\ \bibnamefont {Chuang}}, \bibinfo {author} {\bibfnamefont {N.}~\bibnamefont {Wiebe}}, \bibinfo {author} {\bibfnamefont {A.}~\bibnamefont {Schuckert}},\ and\ \bibinfo {author} {\bibfnamefont {S.~M.}\ \bibnamefont {Girvin}},\ }\href@noop {} {\bibfield  {title} {\emph {\bibinfo {title} {Hybrid Oscillator-Qubit Quantum Processors: Simulating Fermions, Bosons, and Gauge Fields}}}} (\bibinfo {year} {2024}),\ \Eprint {https://arxiv.org/abs/2409.03747}
  {arXiv:2409.03747} \BibitemShut {NoStop}%
\bibitem [{\citenamefont {Elstner}\ and\ \citenamefont {Monien}(1999)}]{Elstner1999}%
  \BibitemOpen
  \bibfield  {author} {\bibinfo {author} {\bibfnamefont {N.}~\bibnamefont {Elstner}}\ and\ \bibinfo {author} {\bibfnamefont {H.}~\bibnamefont {Monien}},\ }\bibfield  {title} {\emph {\bibinfo {title} {Dynamics and thermodynamics of the Bose-Hubbard model}},\ }\href {https://doi.org/10.1103/PhysRevB.59.12184} {\bibfield  {journal} {\bibinfo  {journal} {Phys. Rev. B}\ }\textbf {\bibinfo {volume} {59}},\ \bibinfo {pages} {12184} (\bibinfo {year} {1999})}\BibitemShut {NoStop}%
\bibitem [{\citenamefont {Schollw\"{o}ck}(2011)}]{Schollwoeck2011}%
  \BibitemOpen
  \bibfield  {author} {\bibinfo {author} {\bibfnamefont {U.}~\bibnamefont {Schollw\"{o}ck}},\ }\bibfield  {title} {\emph {\bibinfo {title} {{T}he density-matrix renormalization group in the age of matrix product states}},\ }\href {https://doi.org/10.1016/j.aop.2010.09.012} {\bibfield  {journal} {\bibinfo  {journal} {Ann. Phys.}\ }\textbf {\bibinfo {volume} {326}},\ \bibinfo {pages} {96} (\bibinfo {year} {2011})}\BibitemShut {NoStop}%
\bibitem [{\citenamefont {Or{\'u}s}(2014)}]{Orus2014}%
  \BibitemOpen
  \bibfield  {author} {\bibinfo {author} {\bibfnamefont {R.}~\bibnamefont {Or{\'u}s}},\ }\bibfield  {title} {\emph {\bibinfo {title} {{A} practical introduction to tensor networks: {M}atrix product states and projected entangled pair states}},\ }\href {https://doi.org/10.1016/j.aop.2014.06.013} {\bibfield  {journal} {\bibinfo  {journal} {Ann. Phys.}\ }\textbf {\bibinfo {volume} {349}},\ \bibinfo {pages} {117} (\bibinfo {year} {2014})}\BibitemShut {NoStop}%
\bibitem [{\citenamefont {Pardo}\ \emph {et~al.}(2023)\citenamefont {Pardo}, \citenamefont {Greenberg}, \citenamefont {Fortinsky}, \citenamefont {Katz},\ and\ \citenamefont {Zohar}}]{Pardo2023}%
  \BibitemOpen
  \bibfield  {author} {\bibinfo {author} {\bibfnamefont {G.}~\bibnamefont {Pardo}}, \bibinfo {author} {\bibfnamefont {T.}~\bibnamefont {Greenberg}}, \bibinfo {author} {\bibfnamefont {A.}~\bibnamefont {Fortinsky}}, \bibinfo {author} {\bibfnamefont {N.}~\bibnamefont {Katz}},\ and\ \bibinfo {author} {\bibfnamefont {E.}~\bibnamefont {Zohar}},\ }\bibfield  {title} {\emph {\bibinfo {title} {Resource-efficient quantum simulation of lattice gauge theories in arbitrary dimensions: Solving for Gauss's law and fermion elimination}},\ }\href {https://doi.org/10.1103/PhysRevResearch.5.023077} {\bibfield  {journal} {\bibinfo  {journal} {Phys. Rev. Res.}\ }\textbf {\bibinfo {volume} {5}},\ \bibinfo {pages} {023077} (\bibinfo {year} {2023})}\BibitemShut {NoStop}%
\bibitem [{\citenamefont {Kebric}\ \emph {et~al.}(2021)\citenamefont {Kebric}, \citenamefont {Barbiero}, \citenamefont {Reinmoser}, \citenamefont {Schollw\"ock},\ and\ \citenamefont {Grusdt}}]{Kebric2021}%
  \BibitemOpen
  \bibfield  {author} {\bibinfo {author} {\bibfnamefont {M.}~\bibnamefont {Kebric}}, \bibinfo {author} {\bibfnamefont {L.}~\bibnamefont {Barbiero}}, \bibinfo {author} {\bibfnamefont {C.}~\bibnamefont {Reinmoser}}, \bibinfo {author} {\bibfnamefont {U.}~\bibnamefont {Schollw\"ock}},\ and\ \bibinfo {author} {\bibfnamefont {F.}~\bibnamefont {Grusdt}},\ }\bibfield  {title} {\emph {\bibinfo {title} {Confinement and Mott Transitions of Dynamical Charges in One-Dimensional Lattice Gauge Theories}},\ }\href {https://doi.org/10.1103/PhysRevLett.127.167203} {\bibfield  {journal} {\bibinfo  {journal} {Phys. Rev. Lett.}\ }\textbf {\bibinfo {volume} {127}},\ \bibinfo {pages} {167203} (\bibinfo {year} {2021})}\BibitemShut {NoStop}%
\bibitem [{\citenamefont {MacDonald}\ \emph {et~al.}(1988)\citenamefont {MacDonald}, \citenamefont {Girvin},\ and\ \citenamefont {Yoshioka}}]{macdonald1988}%
  \BibitemOpen
  \bibfield  {author} {\bibinfo {author} {\bibfnamefont {A.~H.}\ \bibnamefont {MacDonald}}, \bibinfo {author} {\bibfnamefont {S.~M.}\ \bibnamefont {Girvin}},\ and\ \bibinfo {author} {\bibfnamefont {D.}~\bibnamefont {Yoshioka}},\ }\bibfield  {title} {\emph {\bibinfo {title} {$\frac{t}{U}$ expansion for the Hubbard model}},\ }\href {https://doi.org/10.1103/PhysRevB.37.9753} {\bibfield  {journal} {\bibinfo  {journal} {Phys. Rev. B}\ }\textbf {\bibinfo {volume} {37}},\ \bibinfo {pages} {9753} (\bibinfo {year} {1988})}\BibitemShut {NoStop}%
\bibitem [{\citenamefont {Jack}\ and\ \citenamefont {Yamashita}(2005)}]{jack2005}%
  \BibitemOpen
  \bibfield  {author} {\bibinfo {author} {\bibfnamefont {M.~W.}\ \bibnamefont {Jack}}\ and\ \bibinfo {author} {\bibfnamefont {M.}~\bibnamefont {Yamashita}},\ }\bibfield  {title} {\emph {\bibinfo {title} {Bose-Hubbard model with attractive interactions}},\ }\href {https://doi.org/10.1103/PhysRevA.71.023610} {\bibfield  {journal} {\bibinfo  {journal} {Phys. Rev. A}\ }\textbf {\bibinfo {volume} {71}},\ \bibinfo {pages} {023610} (\bibinfo {year} {2005})}\BibitemShut {NoStop}%
\bibitem [{\citenamefont {Mansikkam\"aki}\ \emph {et~al.}(2022)\citenamefont {Mansikkam\"aki}, \citenamefont {Laine}, \citenamefont {Piltonen},\ and\ \citenamefont {Silveri}}]{Mansikkamaki2022}%
  \BibitemOpen
  \bibfield  {author} {\bibinfo {author} {\bibfnamefont {O.}~\bibnamefont {Mansikkam\"aki}}, \bibinfo {author} {\bibfnamefont {S.}~\bibnamefont {Laine}}, \bibinfo {author} {\bibfnamefont {A.}~\bibnamefont {Piltonen}},\ and\ \bibinfo {author} {\bibfnamefont {M.}~\bibnamefont {Silveri}},\ }\bibfield  {title} {\emph {\bibinfo {title} {Beyond Hard-Core Bosons in Transmon Arrays}},\ }\href {https://doi.org/10.1103/PRXQuantum.3.040314} {\bibfield  {journal} {\bibinfo  {journal} {PRX Quantum}\ }\textbf {\bibinfo {volume} {3}},\ \bibinfo {pages} {040314} (\bibinfo {year} {2022})}\BibitemShut {NoStop}%
\bibitem [{\citenamefont {Dyson}\ and\ \citenamefont {Lenard}(1967)}]{Dyson1967}%
  \BibitemOpen
  \bibfield  {author} {\bibinfo {author} {\bibfnamefont {F.~J.}\ \bibnamefont {Dyson}}\ and\ \bibinfo {author} {\bibfnamefont {A.}~\bibnamefont {Lenard}},\ }\bibfield  {title} {\emph {\bibinfo {title} {Stability of Matter. I}},\ }\href {https://doi.org/10.1063/1.1705209} {\bibfield  {journal} {\bibinfo  {journal} {J. Math. Phys.}\ }\textbf {\bibinfo {volume} {8}},\ \bibinfo {pages} {423–434} (\bibinfo {year} {1967})}\BibitemShut {NoStop}%
\bibitem [{\citenamefont {Deng}\ \emph {et~al.}(2008)\citenamefont {Deng}, \citenamefont {Porras},\ and\ \citenamefont {Cirac}}]{Deng2008}%
  \BibitemOpen
  \bibfield  {author} {\bibinfo {author} {\bibfnamefont {X.-L.}\ \bibnamefont {Deng}}, \bibinfo {author} {\bibfnamefont {D.}~\bibnamefont {Porras}},\ and\ \bibinfo {author} {\bibfnamefont {J.~I.}\ \bibnamefont {Cirac}},\ }\bibfield  {title} {\emph {\bibinfo {title} {Quantum phases of interacting phonons in ion traps}},\ }\href {https://doi.org/10.1103/PhysRevA.77.033403} {\bibfield  {journal} {\bibinfo  {journal} {Phys. Rev. A}\ }\textbf {\bibinfo {volume} {77}},\ \bibinfo {pages} {033403} (\bibinfo {year} {2008})}\BibitemShut {NoStop}%
\bibitem [{\citenamefont {Kiely}\ and\ \citenamefont {Mueller}(2022)}]{Kiely2022}%
  \BibitemOpen
  \bibfield  {author} {\bibinfo {author} {\bibfnamefont {T.~G.}\ \bibnamefont {Kiely}}\ and\ \bibinfo {author} {\bibfnamefont {E.~J.}\ \bibnamefont {Mueller}},\ }\bibfield  {title} {\emph {\bibinfo {title} {Superfluidity in the One-Dimensional {{Bose-Hubbard}} Model}},\ }\href {https://doi.org/10.1103/PhysRevB.105.134502} {\bibfield  {journal} {\bibinfo  {journal} {Phys. Rev. B}\ }\textbf {\bibinfo {volume} {105}},\ \bibinfo {pages} {134502} (\bibinfo {year} {2022})}\BibitemShut {NoStop}%
\bibitem [{\citenamefont {Giamarchi}(2003)}]{Giamarchi2003}%
  \BibitemOpen
  \bibfield  {author} {\bibinfo {author} {\bibfnamefont {T.}~\bibnamefont {Giamarchi}},\ }\href {https://doi.org/10.1093/acprof:oso/9780198525004.001.0001} {\emph {\bibinfo {title} {Quantum Physics in One Dimension}}}\ (\bibinfo  {publisher} {Oxford University Press},\ \bibinfo {year} {2003})\BibitemShut {NoStop}%
\bibitem [{\citenamefont {Sachdev}\ \emph {et~al.}(2002)\citenamefont {Sachdev}, \citenamefont {Sengupta},\ and\ \citenamefont {Girvin}}]{Sachdev2002}%
  \BibitemOpen
  \bibfield  {author} {\bibinfo {author} {\bibfnamefont {S.}~\bibnamefont {Sachdev}}, \bibinfo {author} {\bibfnamefont {K.}~\bibnamefont {Sengupta}},\ and\ \bibinfo {author} {\bibfnamefont {S.~M.}\ \bibnamefont {Girvin}},\ }\bibfield  {title} {\emph {\bibinfo {title} {Mott insulators in strong electric fields}},\ }\href {https://doi.org/10.1103/PhysRevB.66.075128} {\bibfield  {journal} {\bibinfo  {journal} {Phys. Rev. B}\ }\textbf {\bibinfo {volume} {66}},\ \bibinfo {pages} {075128} (\bibinfo {year} {2002})}\BibitemShut {NoStop}%
\bibitem [{\citenamefont {Guardado-Sanchez}\ \emph {et~al.}(2020)\citenamefont {Guardado-Sanchez}, \citenamefont {Morningstar}, \citenamefont {Spar}, \citenamefont {Brown}, \citenamefont {Huse},\ and\ \citenamefont {Bakr}}]{guardadosanchez2020}%
  \BibitemOpen
  \bibfield  {author} {\bibinfo {author} {\bibfnamefont {E.}~\bibnamefont {Guardado-Sanchez}}, \bibinfo {author} {\bibfnamefont {A.}~\bibnamefont {Morningstar}}, \bibinfo {author} {\bibfnamefont {B.~M.}\ \bibnamefont {Spar}}, \bibinfo {author} {\bibfnamefont {P.~T.}\ \bibnamefont {Brown}}, \bibinfo {author} {\bibfnamefont {D.~A.}\ \bibnamefont {Huse}},\ and\ \bibinfo {author} {\bibfnamefont {W.~S.}\ \bibnamefont {Bakr}},\ }\bibfield  {title} {\emph {\bibinfo {title} {Subdiffusion and Heat Transport in a Tilted Two-Dimensional Fermi-Hubbard System}},\ }\href {https://doi.org/10.1103/PhysRevX.10.011042} {\bibfield  {journal} {\bibinfo  {journal} {Phys. Rev. X}\ }\textbf {\bibinfo {volume} {10}},\ \bibinfo {pages} {011042} (\bibinfo {year} {2020})}\BibitemShut {NoStop}%
\bibitem [{\citenamefont {Zechmann}\ \emph {et~al.}(2023)\citenamefont {Zechmann}, \citenamefont {Altman}, \citenamefont {Knap},\ and\ \citenamefont {Feldmeier}}]{Zechmann2023}%
  \BibitemOpen
  \bibfield  {author} {\bibinfo {author} {\bibfnamefont {P.}~\bibnamefont {Zechmann}}, \bibinfo {author} {\bibfnamefont {E.}~\bibnamefont {Altman}}, \bibinfo {author} {\bibfnamefont {M.}~\bibnamefont {Knap}},\ and\ \bibinfo {author} {\bibfnamefont {J.}~\bibnamefont {Feldmeier}},\ }\bibfield  {title} {\emph {\bibinfo {title} {Fractonic Luttinger liquids and supersolids in a constrained Bose-Hubbard model}},\ }\href {https://doi.org/10.1103/PhysRevB.107.195131} {\bibfield  {journal} {\bibinfo  {journal} {Phys. Rev. B}\ }\textbf {\bibinfo {volume} {107}},\ \bibinfo {pages} {195131} (\bibinfo {year} {2023})}\BibitemShut {NoStop}%
\bibitem [{\citenamefont {Sala}\ \emph {et~al.}(2020)\citenamefont {Sala}, \citenamefont {Rakovszky}, \citenamefont {Verresen}, \citenamefont {Knap},\ and\ \citenamefont {Pollmann}}]{sala2020}%
  \BibitemOpen
  \bibfield  {author} {\bibinfo {author} {\bibfnamefont {P.}~\bibnamefont {Sala}}, \bibinfo {author} {\bibfnamefont {T.}~\bibnamefont {Rakovszky}}, \bibinfo {author} {\bibfnamefont {R.}~\bibnamefont {Verresen}}, \bibinfo {author} {\bibfnamefont {M.}~\bibnamefont {Knap}},\ and\ \bibinfo {author} {\bibfnamefont {F.}~\bibnamefont {Pollmann}},\ }\bibfield  {title} {\emph {\bibinfo {title} {Ergodicity Breaking Arising from Hilbert Space Fragmentation in Dipole-Conserving Hamiltonians}},\ }\href {https://doi.org/10.1103/PhysRevX.10.011047} {\bibfield  {journal} {\bibinfo  {journal} {Phys. Rev. X}\ }\textbf {\bibinfo {volume} {10}},\ \bibinfo {pages} {011047} (\bibinfo {year} {2020})}\BibitemShut {NoStop}%
\bibitem [{\citenamefont {Burchards}\ \emph {et~al.}(2022)\citenamefont {Burchards}, \citenamefont {Feldmeier}, \citenamefont {Schuckert},\ and\ \citenamefont {Knap}}]{Burchards2022}%
  \BibitemOpen
  \bibfield  {author} {\bibinfo {author} {\bibfnamefont {A.~G.}\ \bibnamefont {Burchards}}, \bibinfo {author} {\bibfnamefont {J.}~\bibnamefont {Feldmeier}}, \bibinfo {author} {\bibfnamefont {A.}~\bibnamefont {Schuckert}},\ and\ \bibinfo {author} {\bibfnamefont {M.}~\bibnamefont {Knap}},\ }\bibfield  {title} {\emph {\bibinfo {title} {Coupled hydrodynamics in dipole-conserving quantum systems}},\ }\href {https://doi.org/10.1103/PhysRevB.105.205127} {\bibfield  {journal} {\bibinfo  {journal} {Phys. Rev. B}\ }\textbf {\bibinfo {volume} {105}},\ \bibinfo {pages} {205127} (\bibinfo {year} {2022})}\BibitemShut {NoStop}%
\bibitem [{\citenamefont {Whitlow}\ \emph {et~al.}(2023)\citenamefont {Whitlow}, \citenamefont {Jia}, \citenamefont {Wang}, \citenamefont {Fang}, \citenamefont {Kim},\ and\ \citenamefont {Brown}}]{Whitlow2023}%
  \BibitemOpen
  \bibfield  {author} {\bibinfo {author} {\bibfnamefont {J.}~\bibnamefont {Whitlow}}, \bibinfo {author} {\bibfnamefont {Z.}~\bibnamefont {Jia}}, \bibinfo {author} {\bibfnamefont {Y.}~\bibnamefont {Wang}}, \bibinfo {author} {\bibfnamefont {C.}~\bibnamefont {Fang}}, \bibinfo {author} {\bibfnamefont {J.}~\bibnamefont {Kim}},\ and\ \bibinfo {author} {\bibfnamefont {K.~R.}\ \bibnamefont {Brown}},\ }\bibfield  {title} {\emph {\bibinfo {title} {Quantum simulation of conical intersections using trapped ions}},\ }\href {https://doi.org/10.1038/s41557-023-01303-0} {\bibfield  {journal} {\bibinfo  {journal} {Nat. Chem.}\ }\textbf {\bibinfo {volume} {15}},\ \bibinfo {pages} {1509} (\bibinfo {year} {2023})}\BibitemShut {NoStop}%
\bibitem [{\citenamefont {Wang}\ \emph {et~al.}(2020)\citenamefont {Wang}, \citenamefont {Curtis}, \citenamefont {Lester}, \citenamefont {Zhang}, \citenamefont {Gao}, \citenamefont {Freeze}, \citenamefont {Batista}, \citenamefont {Vaccaro}, \citenamefont {Chuang}, \citenamefont {Frunzio}, \citenamefont {Jiang}, \citenamefont {Girvin},\ and\ \citenamefont {Schoelkopf}}]{Wang2020}%
  \BibitemOpen
  \bibfield  {author} {\bibinfo {author} {\bibfnamefont {C.~S.}\ \bibnamefont {Wang}}, \bibinfo {author} {\bibfnamefont {J.~C.}\ \bibnamefont {Curtis}}, \bibinfo {author} {\bibfnamefont {B.~J.}\ \bibnamefont {Lester}}, \bibinfo {author} {\bibfnamefont {Y.}~\bibnamefont {Zhang}}, \bibinfo {author} {\bibfnamefont {Y.~Y.}\ \bibnamefont {Gao}}, \bibinfo {author} {\bibfnamefont {J.}~\bibnamefont {Freeze}}, \bibinfo {author} {\bibfnamefont {V.~S.}\ \bibnamefont {Batista}}, \bibinfo {author} {\bibfnamefont {P.~H.}\ \bibnamefont {Vaccaro}}, \bibinfo {author} {\bibfnamefont {I.~L.}\ \bibnamefont {Chuang}}, \bibinfo {author} {\bibfnamefont {L.}~\bibnamefont {Frunzio}}, \bibinfo {author} {\bibfnamefont {L.}~\bibnamefont {Jiang}}, \bibinfo {author} {\bibfnamefont {S.~M.}\ \bibnamefont {Girvin}},\ and\ \bibinfo {author} {\bibfnamefont {R.~J.}\ \bibnamefont {Schoelkopf}},\ }\bibfield  {title} {\emph {\bibinfo {title} {Efficient Multiphoton Sampling of Molecular Vibronic Spectra on a Superconducting Bosonic Processor}},\
  }\href {https://doi.org/10.1103/PhysRevX.10.021060} {\bibfield  {journal} {\bibinfo  {journal} {Phys. Rev. X}\ }\textbf {\bibinfo {volume} {10}},\ \bibinfo {pages} {021060} (\bibinfo {year} {2020})}\BibitemShut {NoStop}%
\bibitem [{\citenamefont {Liu}\ \emph {et~al.}(2024)\citenamefont {Liu}, \citenamefont {Singh}, \citenamefont {Smith}, \citenamefont {Crane}, \citenamefont {Martyn}, \citenamefont {Eickbusch}, \citenamefont {Schuckert}, \citenamefont {Li}, \citenamefont {Sinanan-Singh}, \citenamefont {Soley}, \citenamefont {Tsunoda}, \citenamefont {Chuang}, \citenamefont {Wiebe},\ and\ \citenamefont {Girvin}}]{liu2024}%
  \BibitemOpen
  \bibfield  {author} {\bibinfo {author} {\bibfnamefont {Y.}~\bibnamefont {Liu}}, \bibinfo {author} {\bibfnamefont {S.}~\bibnamefont {Singh}}, \bibinfo {author} {\bibfnamefont {K.~C.}\ \bibnamefont {Smith}}, \bibinfo {author} {\bibfnamefont {E.}~\bibnamefont {Crane}}, \bibinfo {author} {\bibfnamefont {J.~M.}\ \bibnamefont {Martyn}}, \bibinfo {author} {\bibfnamefont {A.}~\bibnamefont {Eickbusch}}, \bibinfo {author} {\bibfnamefont {A.}~\bibnamefont {Schuckert}}, \bibinfo {author} {\bibfnamefont {R.~D.}\ \bibnamefont {Li}}, \bibinfo {author} {\bibfnamefont {J.}~\bibnamefont {Sinanan-Singh}}, \bibinfo {author} {\bibfnamefont {M.~B.}\ \bibnamefont {Soley}}, \bibinfo {author} {\bibfnamefont {T.}~\bibnamefont {Tsunoda}}, \bibinfo {author} {\bibfnamefont {I.~L.}\ \bibnamefont {Chuang}}, \bibinfo {author} {\bibfnamefont {N.}~\bibnamefont {Wiebe}},\ and\ \bibinfo {author} {\bibfnamefont {S.~M.}\ \bibnamefont {Girvin}},\ }\href@noop {} {\bibfield  {title} {\emph {\bibinfo {title} {Hybrid Oscillator-Qubit Quantum
  Processors: Instruction Set Architectures, Abstract Machine Models, and Applications}}}} (\bibinfo {year} {2024}),\ \Eprint {https://arxiv.org/abs/2407.10381} {arXiv:2407.10381} \BibitemShut {NoStop}%
\bibitem [{\citenamefont {Kebric}\ \emph {et~al.}(2024)\citenamefont {Kebric}, \citenamefont {Halimeh}, \citenamefont {Schollw\"ock},\ and\ \citenamefont {Grusdt}}]{kebric2023a}%
  \BibitemOpen
  \bibfield  {author} {\bibinfo {author} {\bibfnamefont {M.}~\bibnamefont {Kebric}}, \bibinfo {author} {\bibfnamefont {J.~C.}\ \bibnamefont {Halimeh}}, \bibinfo {author} {\bibfnamefont {U.}~\bibnamefont {Schollw\"ock}},\ and\ \bibinfo {author} {\bibfnamefont {F.}~\bibnamefont {Grusdt}},\ }\bibfield  {title} {\emph {\bibinfo {title} {Confinement in $(1+1)$-dimensional ${\mathbb{Z}}_{2}$ lattice gauge theories at finite temperature}},\ }\href {https://doi.org/10.1103/PhysRevB.109.245110} {\bibfield  {journal} {\bibinfo  {journal} {Phys. Rev. B}\ }\textbf {\bibinfo {volume} {109}},\ \bibinfo {pages} {245110} (\bibinfo {year} {2024})}\BibitemShut {NoStop}%
\bibitem [{\citenamefont {Schuckert}\ \emph {et~al.}(2025)\citenamefont {Schuckert}, \citenamefont {Katz}, \citenamefont {Feng}, \citenamefont {Crane}, \citenamefont {De}, \citenamefont {Hafezi}, \citenamefont {Gorshkov},\ and\ \citenamefont {Monroe}}]{schuckert2025}%
  \BibitemOpen
  \bibfield  {author} {\bibinfo {author} {\bibfnamefont {A.}~\bibnamefont {Schuckert}}, \bibinfo {author} {\bibfnamefont {O.}~\bibnamefont {Katz}}, \bibinfo {author} {\bibfnamefont {L.}~\bibnamefont {Feng}}, \bibinfo {author} {\bibfnamefont {E.}~\bibnamefont {Crane}}, \bibinfo {author} {\bibfnamefont {A.}~\bibnamefont {De}}, \bibinfo {author} {\bibfnamefont {M.}~\bibnamefont {Hafezi}}, \bibinfo {author} {\bibfnamefont {A.~V.}\ \bibnamefont {Gorshkov}},\ and\ \bibinfo {author} {\bibfnamefont {C.}~\bibnamefont {Monroe}},\ }\bibfield  {title} {\emph {\bibinfo {title} {Observation of a Finite-Energy Phase Transition in a One-Dimensional Quantum Simulator}},\ }\href {https://doi.org/10.1038/s41567-024-02751-2} {\bibfield  {journal} {\bibinfo  {journal} {Nat. Phys.}\ } (\bibinfo {year} {2025})}\BibitemShut {NoStop}%
\bibitem [{\citenamefont {H\'emery}\ \emph {et~al.}(2024)\citenamefont {H\'emery}, \citenamefont {Ghanem}, \citenamefont {Crane}, \citenamefont {Campbell}, \citenamefont {Dreiling}, \citenamefont {Figgatt}, \citenamefont {Foltz}, \citenamefont {Gaebler}, \citenamefont {Johansen}, \citenamefont {Mills}, \citenamefont {Moses}, \citenamefont {Pino}, \citenamefont {Ransford}, \citenamefont {Rowe}, \citenamefont {Siegfried}, \citenamefont {Stutz}, \citenamefont {Dreyer}, \citenamefont {Schuckert},\ and\ \citenamefont {Nigmatullin}}]{Hemery2024}%
  \BibitemOpen
  \bibfield  {author} {\bibinfo {author} {\bibfnamefont {K.}~\bibnamefont {H\'emery}}, \bibinfo {author} {\bibfnamefont {K.}~\bibnamefont {Ghanem}}, \bibinfo {author} {\bibfnamefont {E.}~\bibnamefont {Crane}}, \bibinfo {author} {\bibfnamefont {S.~L.}\ \bibnamefont {Campbell}}, \bibinfo {author} {\bibfnamefont {J.~M.}\ \bibnamefont {Dreiling}}, \bibinfo {author} {\bibfnamefont {C.}~\bibnamefont {Figgatt}}, \bibinfo {author} {\bibfnamefont {C.}~\bibnamefont {Foltz}}, \bibinfo {author} {\bibfnamefont {J.~P.}\ \bibnamefont {Gaebler}}, \bibinfo {author} {\bibfnamefont {J.}~\bibnamefont {Johansen}}, \bibinfo {author} {\bibfnamefont {M.}~\bibnamefont {Mills}}, \bibinfo {author} {\bibfnamefont {S.~A.}\ \bibnamefont {Moses}}, \bibinfo {author} {\bibfnamefont {J.~M.}\ \bibnamefont {Pino}}, \bibinfo {author} {\bibfnamefont {A.}~\bibnamefont {Ransford}}, \bibinfo {author} {\bibfnamefont {M.}~\bibnamefont {Rowe}}, \bibinfo {author} {\bibfnamefont {P.}~\bibnamefont {Siegfried}}, \bibinfo {author} {\bibfnamefont {R.~P.}\
  \bibnamefont {Stutz}}, \bibinfo {author} {\bibfnamefont {H.}~\bibnamefont {Dreyer}}, \bibinfo {author} {\bibfnamefont {A.}~\bibnamefont {Schuckert}},\ and\ \bibinfo {author} {\bibfnamefont {R.}~\bibnamefont {Nigmatullin}},\ }\bibfield  {title} {\emph {\bibinfo {title} {Measuring the Loschmidt Amplitude for Finite-Energy Properties of the Fermi-Hubbard Model on an Ion-Trap Quantum Computer}},\ }\href {https://doi.org/10.1103/PRXQuantum.5.030323} {\bibfield  {journal} {\bibinfo  {journal} {PRX Quantum}\ }\textbf {\bibinfo {volume} {5}},\ \bibinfo {pages} {030323} (\bibinfo {year} {2024})}\BibitemShut {NoStop}%
\bibitem [{\citenamefont {Nigmatullin}\ \emph {et~al.}(2024)\citenamefont {Nigmatullin}, \citenamefont {Hemery}, \citenamefont {Ghanem}, \citenamefont {Moses}, \citenamefont {Gresh}, \citenamefont {Siegfried}, \citenamefont {Mills}, \citenamefont {Gatterman}, \citenamefont {Hewitt}, \citenamefont {Granet},\ and\ \citenamefont {Dreyer}}]{nigmatullin2024}%
  \BibitemOpen
  \bibfield  {author} {\bibinfo {author} {\bibfnamefont {R.}~\bibnamefont {Nigmatullin}}, \bibinfo {author} {\bibfnamefont {K.}~\bibnamefont {Hemery}}, \bibinfo {author} {\bibfnamefont {K.}~\bibnamefont {Ghanem}}, \bibinfo {author} {\bibfnamefont {S.}~\bibnamefont {Moses}}, \bibinfo {author} {\bibfnamefont {D.}~\bibnamefont {Gresh}}, \bibinfo {author} {\bibfnamefont {P.}~\bibnamefont {Siegfried}}, \bibinfo {author} {\bibfnamefont {M.}~\bibnamefont {Mills}}, \bibinfo {author} {\bibfnamefont {T.}~\bibnamefont {Gatterman}}, \bibinfo {author} {\bibfnamefont {N.}~\bibnamefont {Hewitt}}, \bibinfo {author} {\bibfnamefont {E.}~\bibnamefont {Granet}},\ and\ \bibinfo {author} {\bibfnamefont {H.}~\bibnamefont {Dreyer}},\ }\href@noop {} {\bibfield  {title} {\emph {\bibinfo {title} {Experimental Demonstration of Break-Even for the Compact Fermionic Encoding}}}} (\bibinfo {year} {2024}),\ \Eprint {https://arxiv.org/abs/2409.06789} {arXiv:2409.06789} \BibitemShut {NoStop}%
\bibitem [{\citenamefont {Evered}\ \emph {et~al.}(2025)\citenamefont {Evered}, \citenamefont {Kalinowski}, \citenamefont {Geim}, \citenamefont {Manovitz}, \citenamefont {Bluvstein}, \citenamefont {Li}, \citenamefont {Maskara}, \citenamefont {Zhou}, \citenamefont {Ebadi}, \citenamefont {Xu}, \citenamefont {Campo}, \citenamefont {Cain}, \citenamefont {Ostermann}, \citenamefont {Yelin}, \citenamefont {Sachdev}, \citenamefont {Greiner}, \citenamefont {Vuletić},\ and\ \citenamefont {Lukin}}]{evered2025}%
  \BibitemOpen
  \bibfield  {author} {\bibinfo {author} {\bibfnamefont {S.~J.}\ \bibnamefont {Evered}}, \bibinfo {author} {\bibfnamefont {M.}~\bibnamefont {Kalinowski}}, \bibinfo {author} {\bibfnamefont {A.~A.}\ \bibnamefont {Geim}}, \bibinfo {author} {\bibfnamefont {T.}~\bibnamefont {Manovitz}}, \bibinfo {author} {\bibfnamefont {D.}~\bibnamefont {Bluvstein}}, \bibinfo {author} {\bibfnamefont {S.~H.}\ \bibnamefont {Li}}, \bibinfo {author} {\bibfnamefont {N.}~\bibnamefont {Maskara}}, \bibinfo {author} {\bibfnamefont {H.}~\bibnamefont {Zhou}}, \bibinfo {author} {\bibfnamefont {S.}~\bibnamefont {Ebadi}}, \bibinfo {author} {\bibfnamefont {M.}~\bibnamefont {Xu}}, \bibinfo {author} {\bibfnamefont {J.}~\bibnamefont {Campo}}, \bibinfo {author} {\bibfnamefont {M.}~\bibnamefont {Cain}}, \bibinfo {author} {\bibfnamefont {S.}~\bibnamefont {Ostermann}}, \bibinfo {author} {\bibfnamefont {S.~F.}\ \bibnamefont {Yelin}}, \bibinfo {author} {\bibfnamefont {S.}~\bibnamefont {Sachdev}}, \bibinfo {author} {\bibfnamefont {M.}~\bibnamefont
  {Greiner}}, \bibinfo {author} {\bibfnamefont {V.}~\bibnamefont {Vuletić}},\ and\ \bibinfo {author} {\bibfnamefont {M.~D.}\ \bibnamefont {Lukin}},\ }\href@noop {} {\bibfield  {title} {\emph {\bibinfo {title} {Probing topological matter and fermion dynamics on a neutral-atom quantum computer}}}} (\bibinfo {year} {2025}),\ \Eprint {https://arxiv.org/abs/2501.18554} {arXiv:2501.18554} \BibitemShut {NoStop}%
\bibitem [{\citenamefont {{Gonz{\'a}lez-Cuadra}}\ \emph {et~al.}(2023)\citenamefont {{Gonz{\'a}lez-Cuadra}}, \citenamefont {Bluvstein}, \citenamefont {Kalinowski}, \citenamefont {Kaubruegger}, \citenamefont {Maskara}, \citenamefont {Naldesi}, \citenamefont {Zache}, \citenamefont {Kaufman}, \citenamefont {Lukin}, \citenamefont {Pichler}, \citenamefont {Vermersch}, \citenamefont {Ye},\ and\ \citenamefont {Zoller}}]{gonzalezcuadra2023}%
  \BibitemOpen
  \bibfield  {author} {\bibinfo {author} {\bibfnamefont {D.}~\bibnamefont {{Gonz{\'a}lez-Cuadra}}}, \bibinfo {author} {\bibfnamefont {D.}~\bibnamefont {Bluvstein}}, \bibinfo {author} {\bibfnamefont {M.}~\bibnamefont {Kalinowski}}, \bibinfo {author} {\bibfnamefont {R.}~\bibnamefont {Kaubruegger}}, \bibinfo {author} {\bibfnamefont {N.}~\bibnamefont {Maskara}}, \bibinfo {author} {\bibfnamefont {P.}~\bibnamefont {Naldesi}}, \bibinfo {author} {\bibfnamefont {T.~V.}\ \bibnamefont {Zache}}, \bibinfo {author} {\bibfnamefont {A.~M.}\ \bibnamefont {Kaufman}}, \bibinfo {author} {\bibfnamefont {M.~D.}\ \bibnamefont {Lukin}}, \bibinfo {author} {\bibfnamefont {H.}~\bibnamefont {Pichler}}, \bibinfo {author} {\bibfnamefont {B.}~\bibnamefont {Vermersch}}, \bibinfo {author} {\bibfnamefont {J.}~\bibnamefont {Ye}},\ and\ \bibinfo {author} {\bibfnamefont {P.}~\bibnamefont {Zoller}},\ }\bibfield  {title} {\emph {\bibinfo {title} {Fermionic Quantum Processing with Programmable Neutral Atom Arrays}},\ }\href
  {https://doi.org/10.1073/pnas.2304294120} {\bibfield  {journal} {\bibinfo  {journal} {PNAS}\ }\textbf {\bibinfo {volume} {120}},\ \bibinfo {pages} {e2304294120} (\bibinfo {year} {2023})}\BibitemShut {NoStop}%
\bibitem [{\citenamefont {Zache}\ \emph {et~al.}(2023)\citenamefont {Zache}, \citenamefont {González-Cuadra},\ and\ \citenamefont {Zoller}}]{zache2023}%
  \BibitemOpen
  \bibfield  {author} {\bibinfo {author} {\bibfnamefont {T.~V.}\ \bibnamefont {Zache}}, \bibinfo {author} {\bibfnamefont {D.}~\bibnamefont {González-Cuadra}},\ and\ \bibinfo {author} {\bibfnamefont {P.}~\bibnamefont {Zoller}},\ }\bibfield  {title} {\emph {\bibinfo {title} {Fermion-qudit quantum processors for simulating lattice gauge theories with matter}},\ }\href {https://doi.org/10.22331/q-2023-10-16-1140} {\bibfield  {journal} {\bibinfo  {journal} {Quantum}\ }\textbf {\bibinfo {volume} {7}},\ \bibinfo {pages} {1140} (\bibinfo {year} {2023})}\BibitemShut {NoStop}%
\bibitem [{\citenamefont {Rad}\ \emph {et~al.}(2024)\citenamefont {Rad}, \citenamefont {Schuckert}, \citenamefont {Crane}, \citenamefont {Nambiar}, \citenamefont {Fei}, \citenamefont {Wyrick}, \citenamefont {Silver}, \citenamefont {Hafezi}, \citenamefont {Davoudi},\ and\ \citenamefont {Gullans}}]{rad2024}%
  \BibitemOpen
  \bibfield  {author} {\bibinfo {author} {\bibfnamefont {A.}~\bibnamefont {Rad}}, \bibinfo {author} {\bibfnamefont {A.}~\bibnamefont {Schuckert}}, \bibinfo {author} {\bibfnamefont {E.}~\bibnamefont {Crane}}, \bibinfo {author} {\bibfnamefont {G.}~\bibnamefont {Nambiar}}, \bibinfo {author} {\bibfnamefont {F.}~\bibnamefont {Fei}}, \bibinfo {author} {\bibfnamefont {J.}~\bibnamefont {Wyrick}}, \bibinfo {author} {\bibfnamefont {R.~M.}\ \bibnamefont {Silver}}, \bibinfo {author} {\bibfnamefont {M.}~\bibnamefont {Hafezi}}, \bibinfo {author} {\bibfnamefont {Z.}~\bibnamefont {Davoudi}},\ and\ \bibinfo {author} {\bibfnamefont {M.~J.}\ \bibnamefont {Gullans}},\ }\href@noop {} {\bibfield  {title} {\emph {\bibinfo {title} {Analog Quantum Simulator of a Quantum Field Theory with Fermion-Spin Systems in Silicon}}}} (\bibinfo {year} {2024}),\ \Eprint {https://arxiv.org/abs/2407.03419} {arXiv:2407.03419} \BibitemShut {NoStop}%
\bibitem [{\citenamefont {Schuckert}\ \emph {et~al.}(2024)\citenamefont {Schuckert}, \citenamefont {Crane}, \citenamefont {Gorshkov}, \citenamefont {Hafezi},\ and\ \citenamefont {Gullans}}]{schuckert2024}%
  \BibitemOpen
  \bibfield  {author} {\bibinfo {author} {\bibfnamefont {A.}~\bibnamefont {Schuckert}}, \bibinfo {author} {\bibfnamefont {E.}~\bibnamefont {Crane}}, \bibinfo {author} {\bibfnamefont {A.~V.}\ \bibnamefont {Gorshkov}}, \bibinfo {author} {\bibfnamefont {M.}~\bibnamefont {Hafezi}},\ and\ \bibinfo {author} {\bibfnamefont {M.~J.}\ \bibnamefont {Gullans}},\ }\href@noop {} {\bibfield  {title} {\emph {\bibinfo {title} {Fermion-Qubit Fault-Tolerant Quantum Computing}}}} (\bibinfo {year} {2024}),\ \Eprint {https://arxiv.org/abs/2411.08955} {arXiv:2411.08955} \BibitemShut {NoStop}%
\end{thebibliography}%

\section*{End Matter}

\textbf{Schrieffer-Wolff transformation.} We derive Eq.~\ref{eq:effH}. For now, consider $U=0$. We label the electric field term $\hat{H_0}$, and the hopping term $\hat{V}$. In the Schrieffer-Wolff procedure, the Hamiltonian is transformed under a unitary $e^{\hat S}$, $\hat H_\mathrm{eff}\equiv e^{\hat S}\hat He^{-\hat S}$, with $\|\hat S\|$ small such that
\begin{align*}
	 \hat H_\mathrm{eff}\approx \hat H_0 +\hat V +[\hat S,\hat H_0] + [\hat S,\hat V] +\frac{1}{2} [\hat S,[\hat S,\hat H_0]] +\dots. ~\label{eq:effH}
\end{align*}
Our goal is to find $\hat S$ such that $[\hat S,\hat H_0]=-\hat V$ leading to $\hat H_\mathrm{eff}\approx \hat H_0 + \frac{1}{2} [\hat S,\hat V]$. By inspection, we find that
\begin{equation*}
	\hat S=(-i)\frac{J}{2g} \sum_i (\hat a_i^\dagger \hat a_{i+1}+h.c.)\hat Y_{i,i+1}
\end{equation*}
fulfills this condition. Inserting $\hat S$ into $\hat H_\mathrm{eff}$, we find Eq.~\ref{eq:effH} for $U=0$. 

We now check whether for $U\neq 0$, the same transformation can be used. The additional relevant term in the expansion in Eq.~\eqref{eq:effH} is $[\hat S, \frac{U}{2}\sum_i \hat n_i^2]=\mathcal{O}(U\frac{J}{g})$. Therefore, we can neglect this term if $\frac{UJ}{g}\ll \frac{J^2}{g} \leftrightarrow \frac{U}{J}\ll 1$. Combining this condition with the condition $\frac{J}{g}\ll 1$ we find that the Hamiltonian in Eq.~\ref{eq:effH} is valid in the regime
\begin{equation*}
    U\ll J \ll g.
\end{equation*}
Importantly, this condition can be fulfilled while $U>\frac{J^2}{g}$, which means that the bunching-superfluid transition is captured by this perturbative Hamiltonian.

\textbf{Stability of matter.} The question about the stability of matter~\cite{Dyson1967} is best formulated in a grand-canonical setting in which there is an infinite particle reservoir. Stability means that the ground state of the system has a finite energy and a finite number of particles, or in other words, that there is a finite number of particles $N$ at which the energy difference between the cost $\mu$ of creating another particle (or a pair of particles) and the interaction energy is positive. If the energy is negative and superextensive, i.e. $E\propto -N^{1+\alpha}$, $\alpha>0$, then there is always an $N$ such that $\mu-E/N<0$ for all $\mu$. Therefore, the ground state of such a system would be a state with infinitely many particles and an infinite amount of energy would have been absorbed by the environment. Therefore, if the energy is negative, then it must be extensive, i.e. $\alpha=0$, because in that case, there is a $\mu$ such that $\mu-E/N>0$, meaning that the chemical potential does what we expect it to: determine how many particles there are in the system.

\begin{figure}[t!]
    \centering
    \includegraphics{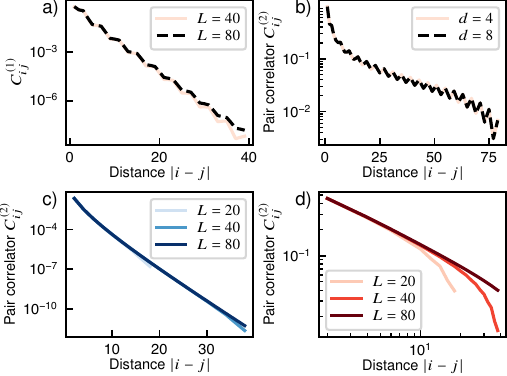}
    \caption{\textbf{Convergence of DMRG correlation functions for $U=50$.} a) Gauge-invariant correlator for different system sizes. b) Pair correlator for different on-site Hilbert-space dimension $d$ for $g/J=28$. $d$ is defined to include the vacuum state, i.e. $d=4$ indicates that up to and including three bosons can occupy each site. c) Pair correlator for different system sizes in the Mott-insulating phase, $g/J=24$. d) Pair correlator for different system sizes in the superfluid phase, $g/J=28$.}
    \label{fig:supp_checks}
\end{figure}

\textbf{DMRG convergence of correlation functions.} We show that the correlation functions shown in Fig.~\ref{fig:SFMI}b-c are converged with respect to system size in Fig.~\ref{fig:supp_checks}, where we only show the converged distance range up to $|i-j|<30$.

\textbf{Parity of a lattice condensate.} Our goal is to calculate the parity $\braket{e^{i\pi \hat n_j}}$ of a condensate, given by the wavefunction $ \ket{\psi} = \ket{\alpha} = e^{-\alpha \hat a^\dagger_{k=0}} \ket{\mathrm{vacuum}}$,
where $\hat a^\dagger_{k} = \frac{1}{\sqrt{L}} \sum_j e^{-ikj} \hat a^\dagger_j$. We also specialize to the case of uniform unit filling, i.e. $\braket{\hat n_j} = 1$ such that $|\alpha|^2=L$, where $L$ is the number of sites.

To calculate the parity, we need to calculate all moments $\braket{(\hat n_j)^m}$. To do so, we expand the density into creation and annihilation operators, $\hat n_j = \frac{1}{L} \sum_{k,k'} e^{i(k-k')j} \hat a_k^\dagger \hat a_{k'}$,
such that
\begin{widetext}
\begin{align}
    \braket{(\hat n_j)^m} &= \frac{1}{L^m} \sum_{k_0,\dots,k_{2m-1}} e^{ij\sum_{l=0}^{2m-1} (k_{2l}-k_{2l+1})} \braket{ \hat a_{k_0}^\dagger \hat a_{k_1} \cdots \hat a_{k_{2m-2}}^\dagger \hat a_{k_{2m-1}}} \\
    &= \frac{1}{L^{m-1}} \sum_{k_1,\dots,k_{2m-2}} e^{ij(k_1+k_{2m-2})+ij\sum_{l=1}^{2m-2} (k_{2l}-k_{2l+1})} \braket{\hat a_{k_1} \cdots \hat a_{k_{2m-2}}^\dagger },
\end{align}
\end{widetext}
where we used the properties of the coherent state $\ket{\psi}$ in the second step. In order to evaluate the remaining expectation value, we rewrite each pair of consecutive annihilation and creation operators as $\hat a_{k_1} \hat a_{k_2}^\dagger = \delta_{k_1,k_2} +  \hat a_{k_2}^\dagger\hat a_{k_1}$. Imagining for the moment that we multiply out all $(m-1)$ terms appearing from this rewriting, we note that each delta function leads to a factor of $L$. Moreover, after complete normal ordering, all momentum sums will disappear such that we can drop the momentum labels on all creation and annihilation operators. Hence, we write 
\begin{align*}
\braket{(\hat n_j)^m} &=\frac{1}{L^{m-1}} 
    \braket{(L+\hat a^\dagger \hat a)^{m-1}}\\
    &= \frac{1}{L^{m-1}} \sum_{l=0}^{m-1} {{m-1}\choose{l}} L^{m-1-l} \braket{(\hat a^\dagger \hat a)^l}\\
    &= \sum_{l=0}^{m-1} {{m-1}\choose{l}}  \braket{(\hat n_j)^l}.
\end{align*}
We therefore found a recursive evaluation of the higher order moments of the occupation number in terms of the lower moments. Comparing this expression with the expression of moments in terms of cumulants $\kappa_m$, 
\begin{equation*}
    \braket{(\hat n_j)^m} = \sum_{l=0}^{m-1} {{m-1}\choose{l}} \kappa_l \braket{(\hat n_j)^l} +\kappa_m,
\end{equation*}
we find $\kappa_m=1$ for all $m$. Finally, the parity with $\pi\rightarrow \alpha$ is in fact the characteristic function of $\hat n_j$, such that it fulfills the identity
\begin{equation*}
    \sum_{k=1}^\infty \frac{1}{k!} (i\alpha)^k \kappa_k = \log\left( \braket{e^{i\alpha \hat n_j}}\right)
\end{equation*}
and therefore $\left( \braket{e^{i\alpha \hat n_j}}\right) = \exp(\exp(i\alpha)-1)$.
This enables us to explicitly calculate all moments and in particular our end result 
\begin{equation*}
    \braket{e^{i\pi \hat n_j}} = \frac{1}{e^2}.
\end{equation*}

By explicitly calculating the moments, one can also convince oneself that a condensate with fixed particle number $\ket{\psi}=\left(\hat a^\dagger_{k=0}\right)^L\ket{\mathrm{vacuum}}$ has the same parity, albeit with strong finite-size corrections.

\end{document}